\documentclass[3p]{elsarticle}

\usepackage{amsmath}
\usepackage{amsfonts}
\usepackage{amssymb}
\usepackage{textcomp}
\usepackage{color}
\usepackage{xcolor}
\usepackage{amsthm}
\usepackage{bm}
\usepackage{mathrsfs}

\usepackage{algorithmicx}
\usepackage{algorithm,algpseudocode}
\usepackage{algpseudocode}
\usepackage{soul}
\usepackage{graphicx}
\usepackage{subcaption}
\usepackage{epsf}
\usepackage{xpatch}
\usepackage{setspace} 

\usepackage{multirow,booktabs}
\usepackage{latexsym}
\usepackage{hyperref}
\usepackage{cleveref}
\usepackage[export]{adjustbox}
\usepackage{dsfont}
\usepackage{mathcomSTEv4}
\usepackage{tikz}
\usepackage{circuitikz}
\usetikzlibrary{arrows,shapes}
\usepackage{tkz-euclide}
\usetikzlibrary{backgrounds}
\usepackage{mathtools}

\usepackage{epstopdf} 
\DeclareGraphicsExtensions{.jpg,pdf,.png}
\RequirePackage{amsopn}
\usepackage{pgfplots}
\pgfplotsset{width=10cm,compat=1.9,every tick label/.append style={scale=0.5}, every axis label/.append style={scale=0.7}}
\usepackage{array}

\newtheorem{theorem}{\textbf{Theorem}}
\newtheorem{lemma}{\textbf{Lemma}}
\newtheorem{definition}{\textbf{Definition}}
\newtheorem{remark}{\textbf{Remark}}
\newtheorem{corollary}{\textbf{Corollary}}

\xpatchcmd{\proof}{\itshape}{\prooflabelfont}{}{}
\newcommand{\prooflabelfont}{\bfseries}

\DeclareGraphicsExtensions{.jpg,pdf,.png,.bmp}

\usetikzlibrary{positioning}
\usetikzlibrary{shapes,arrows}
\tikzstyle{block} = [draw, fill=white, rectangle, 
minimum height=2em, minimum width=3em]
\tikzstyle{input} = [coordinate]
\tikzstyle{output} = [coordinate]
\tikzstyle{pinstyle} = [pin edge={to-,t,black}]

\newcommand{\Lif}{\mathscr{L}}
\def\change{\color{black}}
\def\changeB{\color{black}}

\definecolor{ao}{rgb}{0.0, 0.0, 0.0}
\def\changeA{\color{black}}

\newcommand{\rom}[1]{\uppercase\expandafter{\romannumeral #1\relax}}

\begin{document}

\doublespacing
\changeB{
\title{Harmonic Retrieval Using Weighted Lifted-Structure Low-Rank Matrix Completion}
}

\begin{frontmatter}

\affiliation[1]{organization={Electrical Engineering Department, Sharif University of Technology}, 
    country={Iran}} 
\affiliation[2]{organization={Department of Electronic Systems, Aalborg university}, 
    country={Denmark}}
\affiliation[3]{organization={School of Electrical Engineering and Computer Science, KTH university}, 
    country={Sweden}} 
\affiliation[4]{organization={Electrical and Computer Engineering Department, National Yang-Ming Chao-Tung  University (NYCU)}, 
    country={Taiwan}}

\author[1,2]{Mohammad~Bokaei}
\author[1,3]{Saeed~Razavikia}
\author[4]{Stefano~Rini}
\author[1]{Arash~Amini}
\author[1]{Hamid~Behrouzi}



\begin{abstract}
{\changeA
In this paper, we investigate the problem of recovering the frequency components of a mixture of $K$ complex sinusoids from a random subset of $N$ equally-spaced time-domain samples. Because of the random subset, the samples are effectively non-uniform. 
Besides, the frequency values of each of the $K$ complex sinusoids are assumed to vary continuously within a given range.
 For this problem, we propose a two-step strategy: (i) we first lift the incomplete set of uniform samples (unavailable samples are treated as missing data) into a structured matrix with missing entries, which is potentially low-rank; then  (ii) we complete the matrix using a weighted nuclear minimization problem. We call the method a \emph{ weighted lifted-structured (WLi) low-rank matrix recovery}. Our approach  can be applied to a range of matrix structures such as Hankel and double-Hankel, among others, and provides improvement over the unweighted existing schemes such as EMaC and DEMaC.}  
{\changeA
We provide theoretical guarantees for the proposed method, as well as numerical simulations in both noiseless and noisy settings. 
Both the theoretical and the numerical results confirm the superiority of the proposed approach.
}

\end{abstract}

\end{frontmatter}




\section{Introduction}
{\changeA
Mixture of complex exponential functions are observed in many real-world applications such as medical imaging \cite{zhang2015accelerating}, astronomical imaging \cite{pan2016towards}, millimeter-wave imaging \cite{sheen2001three},  and artworks \cite{fatemi2016shapes,razavikia2019reconstruction,razavikia2019sampling}. 
For instance, in many imaging systems, the optical point spread function 
 can be fairly approximated by a mixture of a few exponential functions. Therefore, the measurements observed in an imaging system are mainly a combination of a few exponential components. Even if these mixtures are of fairly high dimension, their representation in the Fourier domain consists of a small number of spikes. The spectral estimation challenge is to decompose such a mixture into exponential elements or, simply, to separate the Fourier spikes. As we observe analog signals through digital sensing devices, the sampling resolution is a limiting factor in properly detecting and separating exponential components in a mixture. There is extensive literature within the signal processing community for surpassing the limits imposed on the physical resolution using processing techniques, which are generally known as super-resolution algorithms \cite{krim1996btwo}. 
The technique proposed in \cite{ candes2014towards}  is a famous grid-less example that relies on sparse recovery and compressed sensing tools. 
In this work, equally-spaced samples are employed to separate the components using a convex optimization in the line spectral estimation problem. 
The authors of \cite{tang2013compressed} consider a similar scenario but with random samples. 
For this setting, a probabilistic guarantee for the number of required samples for perfect recovery is provided. 
For perfect recovery, both \cite{ candes2014towards} and \cite{tang2013compressed} require a minimum phase separation between the complex sinusoid components (minimum distance between the spikes). 
Inspired by the matrix pencil algorithm in \cite{hua1992estimating}, \cite{chen2014robust} proposed the EMaC method using matrix completion to reduce this minimum required separation.
The method uses a Hankel lifting structure that transforms uniform samples of exponential mixtures into a low-rank matrix.
Once lifted, the recovery of missing samples can be treated as completing a low-rank matrix, which is a well-studied problem \cite{candes2009exact}.
This approach is further extended in \cite{ye2016compressive} for the larger class of signals with a finite rate of innovation. In \cite{cai2016robust}, the problem of low-rank Hankel matrix recovery for random Gaussian projections is investigated, and a lower bound for sample complexity  (with high probability) is derived. Besides the low-rank property of the matrix, most matrix completion techniques require the available entries to be uniformly spread within the matrix. For the case of non-uniform samples, a two-phase sampling-recovery strategy is proposed in \cite{chen2015completing}. However, the method does not work for structured matrices (such as Hankel matrices). 
}

\noindent
\emph{Contributions:}
\noindent
{\changeA
In this work, we consider the problem of separating exponential signals from a mixture of non-uniform samples. The proposed spectral estimation approach is comprised of two steps. 
We first lift the samples/measurements to a structured low-rank matrix with missing samples. Next, we propose 
a general approach to enhance matrix completion by using a measurement-adaptive weighting scheme in which the weights reflect the relative significance of the samples.
}
 The lifting operation encompasses Hankel, double-Hankel, wrap-around Hankel, Hankel-block-Hankel, Toeplitz, and multi-level Toeplitz structures as special cases. 
While our approach bears similarity to the atomic norm minimizations (ANM)  in  \cite{li2015off,yang2015gridless}, we do not require the statistics of the sources. 
{\changeB Instead, we take advantage of the the concept of  \emph{leverage scores} studied in \cite{chen2015completing} for determining the sample weights based on their informativeness.}
We refer to this approach as ``weighted lifted-structure (WLi) low-rank matrix recovery''. 
We show that the  weighting scheme of \cite{chen2015completing} reduces the number of measurements required for  estimating the exponential components. 
After having presented the weighted lifted-structured low-rank matrix completion in some generality, we consider Hankel and double-Hankel structures as special cases (similar to EMaC \cite{chen2014robust}, and DEMac \cite{yang2021new}, respectively), resulting in  WLi-EMaC and WLi-DEMaC methods. 
The simulation results show that the weighted methods require much fewer samples to recover the input vector than the unweighted scheme.

\noindent
{\bf{Note well:}} This paper is the first part of a two-part submission. In this part, we mainly discuss the theoretical perspective of the weighted recovery in a lifted-structure completion problem. In Part II \cite{partII}, we focus on the specific application of DOA estimation using the theoretical results derived here.
 
\noindent
\emph{Organization:}
\noindent
in Section \ref{sec:Lifting operator}, we introduce lifting operator. The problem formulation is provided in Section \ref{Sec:Model}.  The weighted matrix completion problem is described in Section \ref{Sec:DOAestimation}.
In Section \ref{sec:Theoritical}, we present the theoretical guarantees for the weighted  approach.
Numerical  simulations are provided in Section \ref{Sec:Simulation}. Finally,  we conclude the paper in Section \ref{Sec:Conclude}.  
\noindent
\emph{Notations:}
\noindent
{ \changeA 
We use  lowercase letters, lower and upper-case boldface letters to represent scalars, vectors, and matrices respectively. We further show linear operators and their adjoints  by calligraphic notations such as $\mathcal{X}$ and $\mathcal{X}^{\dagger}$, where the superscript in the latter stands for the adjoint operator. Moreover, $\mathbf{X}^{\mathsf{T}}$ and $\mathbf{X}^{\mathsf{H}}$ denote the transpose and Hermitian of a matrix $\mathbf{X}$, respectively. $\mathbf{X} {\odot} \mathbf{Y}$  and $\langle \mathbf{X}, \mathbf{Y}\rangle$ show  the Hadamard (element-wise) product  and inner product of two equi-size matrices $\mathbf{X}$ and $\mathbf{Y}$, respectively. We denote the spectral, Frobenius, and nuclear norms of a matrix $\mathbf{X}$ by $\|\mathbf{X}\|$, $\|\mathbf{X}\|_{\rm F}$ and $\|\mathbf{X}\|_{\rm *}$, respectively. 
Similarly, $\|\mathbf{X}\|_1$ and $\|\mathbf{X}\|_{\infty}$ stand for the element-wise $\ell_1$ and $\ell_{\infty}$ norms of $\mathbf{X}$ (treating $\mathbf{X}$ as a vector). $\|\mathbf{X}\|_{\infty \rightarrow \infty}$ is defined as ${\max_{i\in[N]} \sum_{j}|x_{i,j}|}$ where $x_{i,j}$ denotes $(i,j)$ element of matrix $\mathbf{X}$. Further,  $\|\mathbf{X}\|_{0}$ is the number of non-zero elements of matrix $\mathbf{X}$. We refer to $\mathbf{e}_i^N$ as the $i$-th canonical  basis vector in dimensional $N$. For an integer $n$, $[n]$ stands for $\{1,2,\dots, n\}$.  We also define the operator $ {\rm diag}: \mathbb{C}^N \mapsto \mathbb{C}^{N\times N}$ maps a vector $\mathbf{x} \in \mathbb{C}^{N}$ into a diagonal matrix with diagonal entries as in $\mathbf{X}$. 
The Greek letter $\Omega$ always reflects a finite set of the integers and $|\Omega|$ denotes its cardinality.}
\section{Lifting operator}
\label{sec:Lifting operator}

Fundamental to our  approach is the  transformation of a vector into a matrix to gain degrees of freedom. We call this transformation the \emph{lifting operator}. 
To properly define the class of lifting operators considered in this paper, we initially introduce the concept of the \emph{lifting basis}.

\begin{definition}
\label{Def:LiftingBasis}
We call $\{\mathbf{A}_n\}_{n \in [N]} \subseteq \mathbb{C}^{d_1 \times d_2}$ a \emph{lifting basis} if 
\begin{enumerate}
    \item for all $1\leq n\leq N$ we have that $\|\mathbf{A}_n\|_{\rm F}=1$,
    
    \item all the non-zero elements of $\mathbf{A}_n$ are positive, real and equal, and
    
    \item  $\mathbf{A}_n$s are orthogonal:
    \begin{align}
        \langle \mathbf{A}_{n_1} \,,\, \mathbf{A}_{n_2} \rangle = {\rm tr}\big(\mathbf{A}_{n_1}^\mathsf{T} \,\mathbf{A}_{n_2}\big) = \delta[n_1-n_2],
    \end{align}
     \item and each column of  $\mathbf{A}_n$ (for $n \in [N]$) has at most one nonzero element, i.e.,
    \begin{align}
         \sum_{j\in[d_2]}\bigg(\sum_{i\in [d_1]}[\mathbf{A}_n]_{i,j}\bigg)^2 =1.
     \end{align}
\end{enumerate}
\end{definition}

For a lifting basis $\{\mathbf{A}_n\}_{n \in [N]}$, Definition \ref{Def:LiftingBasis} implies that the non-zero elements in $\mathbf{A}_n$ are all equal to $\frac{1}{\sqrt{\|\mathbf{A}_n\|_0}}$. This further shows that
\begin{align}
\|\mathbf{A}_n\| \leq \|\mathbf{A}_n\|_{\rm F} = 1.
\end{align}

\begin{definition}
\label{Def:Lifting}
Let $\{\mathbf{A}_n\}_{n \in [N]} \subseteq \mathbb{C}^{d_1 \times d_2}$ be a lifting basis according to Definition \ref{Def:LiftingBasis}. The linear mapping $\mathscr{L}: \mathbb{C}^{N} \mapsto \mathbb{C}^{d_1 \times d_2}$ defined by
\begin{align}
\label{eq:DefLif}
\Lif(\mathbf{x})  = 
\sum_{n\in [N]}
a_n \, \langle\mathbf{e}^N_n,\mathbf{x}\rangle\, \mathbf{A}_{n},
\end{align}
is called a \emph{lifting operator}, where $\{a_n\}_{n \in [N]}\subseteq \mathbb{C}$ are constants. We can  check that
$\Lif^{\dagger} :\mathbb{C}^{d_1\times d_2}\mapsto \mathbb{C}^{N}$ with 
\begin{align}
   \mathbf{M}\in \mathbb{C}^{d_1\times d_2}:~~~  \Lif^{\dagger}(\mathbf{M}) = \sum_{\substack{n\in [N] \\ a_n\neq 0}} \tfrac{1}{a_n}\langle \mathbf{A}_n, \mathbf{M} \rangle \,  \mathbf{e}^N_n,
\end{align}
defines the orthogonal back projection from $\mathbb{C}^{d_1\times d_2}$ into $\mathbb{C}^N$.
\end{definition}

By tuning the lifting basis, one can achieve various matrix structures in the output of the lifting operator, such as Hankel, double-Hankel, wrap-around Hankel, Hankel-block-Hankel, Toeplitz, and multi-level Toeplitz. As an example, we examine the Hankel lifting operator $\mathscr{H}:\mathbb{C}^N \to \mathbb{C}^{d\times (N-d+1)}$ with $d\in[N]$:
\begin{equation}
\label{eq:DefHank}
\mathbf{x} = \left[ \begin{matrix}
x_1\\
x_2\\
\vdots \\
x_N
\end{matrix}\right] ~~ \Rightarrow ~~
\mathscr{H}(\mathbf{x}) := \left[ \begin{matrix}
x_{1} & x_{2}  & \dots& x_{N-d+1} \\
x_{2} & x_{3}  & \dots& x_{N-d+2} \\
\vdots & \vdots & \ddots&  \vdots\\
x_{d} & x_{d+1}  & \dots & x_{N} 
\end{matrix}  \right].
\end{equation}
It is not difficult to
verify that this operator corresponds to 
\begin{align}\label{eq:Ak}
\begin{array}{lll}
1\leq n\leq d: & 
{\mathbf{A}}_{n}^{(\mathscr{H})} :=\frac{\sum\limits_{i\in [n] }{\rm \mathbf{e}}_i^d{{\rm \mathbf{e}}_{n-i+1}^{(N-d+1)}}^{\mathsf{T} }}{\sqrt{n}}, & 
a_{n}^{(\mathscr{H})} := \sqrt{n}, \\
d \hspace{-1mm}+\hspace{-1mm} 1\leq n\leq N \hspace{-1mm}-\hspace{-1mm} d \hspace{-1mm}+\hspace{-1mm} 1: &
 {\mathbf{A}}_{n}^{(\mathscr{H})} := \frac{\sum\limits_{i \in [d]} {\rm \mathbf{e}}_i^d{{\rm \mathbf{e}}_{n-i+1}^{(N-d+1)}}^{\mathsf{T} }}{\sqrt{d}}, & 
 a_{n}^{(\mathscr{H})} := \sqrt{d}, \phantom{\bigg|}\\
N \hspace{-1mm}-\hspace{-1mm} d \hspace{-1mm}+\hspace{-1mm} 2\leq n\leq N: & {\mathbf{A}}_{n}^{(\mathscr{H})} := \frac{  \sum\limits_{i=n-N+d}^{d}{\rm \mathbf{e}}_i^{d}{{\rm \mathbf{e}}_{n-i+1}^{(N-d+1)}}^{\mathsf{T} }}{\sqrt{N-n+1}}, & a_{n}^{(\mathscr{H})} := \sqrt{N \hspace{-1mm}-\hspace{-1mm} n \hspace{-1mm}+\hspace{-1mm} 1}.
\end{array}
\end{align}
Our recovery method uses certain values associated with a lifted structure known as \emph{leverage scores}. These scores were originally defined in \cite{chen2015completing} for an adaptive sampling scheme: if we have observed an incomplete matrix and wish to take some more samples before starting to estimate the unobserved entries, which elements are the best options that facilitate the estimation task. For this purpose, each matrix element was assigned a score in \cite{chen2015completing}, which was later interpreted as the sampling probability; i.e., an unobserved entry with a larger leverage score is more likely to be sampled. It is shown in \cite{chen2015completing} that for a given recovery quality, this strategy requires fewer samples compared to the case of observing the matrix entries uniformly at random. 
In the matrix completion problem of \cite{chen2015completing}, each of the elements of the low-rank matrix $\mathbf{M} \in \mathbb{C}^{d_1 \times d_2}$ could be observed independently of other elements; hence, $d_1 \times d_2$ leverage scores are defined. In contrast, among $d_1\times d_2$ elements of  $\Lif(\mathbf{x})$ only $N$ are different ($\mathbf{x}\in\mathbb{C}^{N}$). This means that $N$ different leverage scores are possible. 
In Definition \ref{LEV_SCORS}, we generalize the concept of leverage scores to the case of lifted structures with reduced degrees of freedom.

\begin{definition}
	\label{LEV_SCORS}
	Let $\Lif$ be a lifting operator with basis $\{\mathbf{A}_n\}_{n \in [N]}$. For each $\mathbf{x}\in\mathbb{C}^N$ we define leverage scores $\{\mu_n\}_{n \in [N]}$ as
	\begin{align}
\label{eq:muDefinition}
	\mu_{n} :=\frac{N}{\tilde{K} } \max\big\{\|\mathbf{U}^{\mathsf{H}}\mathbf{A}_{n}\|_{\rm F}^2,
	 \| \mathbf{A}_{n}\mathbf{V}^{\mathsf{H}} \|_{\rm F}^2   \big\},
	\end{align}
	where $\tilde{K}$ is the rank of $\Lif(\mathbf{x})$ and $\mathbf{U}_{d_1\times \tilde{K}}\boldsymbol{\Sigma}_{\tilde{K}\times \tilde{K}} (\mathbf{V}_{d_2\times \tilde{K}})^{\mathsf{H}}$ represents the singular value decomposition (SVD) of $\Lif(\mathbf{x})$.
\end{definition}

For the particular case of low-rank Hankel matrix completion, it is shown in  \cite[Theorem 1]{chen2014robust} that nuclear-norm minimization succeeds in recovering the matrix if the number of [random] samples exceeds a threshold that is proportional to $\max_{n}\{\mu_{n}\}$. In this paper, we provide recovery guarantees  (noiseless and noisy cases) for sample sizes that scale with $\frac{\sum_{n}\mu_{n}}{N}$.

\section{Signal Model}
\label{Sec:Model}

{\changeA 

Let the signal of interest $y(t)$ be a linear mixture of $K$ exponential components. The samples of this signal (whether available or unavailable) are 
\begin{align}
\label{eq:measured_Discrete}
y_n = y(n)  =\sum_{k \in [K]} b_{k} z_{k}^{n}, \quad n  \in [N] ,
\end{align}
where $\{b_{k}\}_{k \in [K]} \in \Cbb$ are the coefficients in the linear mixture and $\{z_k\}_{k \in [K]}\in \mathbb{C}$ are the complex basis. Using vectorial notations, we write $\mathbf{y}= [ y_1,\dots,y_N]^T \in \mathbb{C}^{N}$. In case we have measurement noise, we have access to the noisy samples (if available)
\begin{align}
\label{eq:noisy_measurement}
\tilde{y}_n = y_n + e_n, 
\end{align}
where $\tilde{\mathbf{y}}=[\tilde{y}_1,\dots,\tilde{y}_N]^T$  and $\mathbf{e}=[e_1,\dots,e_N]^T \in \mathbb{C}^{N}$ stand for the measurement and the noise vectors, respectively.  We assume that for each $n\in[N]$, noise amplitudes are upper-bounded as $|e_n|<\eta$ with high probability.  This implies that $\|\mathbf{e}\|_{2} \leq \sqrt{N} \eta$ (with high probability). 
Further, let $\Omega \subseteq [N]$ with $|\Omega|=M\leq N$ be the index set of available samples, i.e., $y_n$ or $\tilde{y}_n$ is available only if $n\in \Omega$. Mathematically, we denote the vector of available samples as
\begin{align}
\label{eq:Projection}
\mathbf{y}_{\Omega} =\mathcal{P}_{\Omega}(\mathbf{y}), ~\rm{or}~ \tilde{\mathbf{y}}_{\Omega} =\mathcal{P}_{\Omega}(\tilde{\mathbf{y}}),
\end{align} 
for the noiseless and noisy cases, respectively. Here,  $\mathcal{P}_{\Omega}:\mathbb{C}^{N}\to\mathbb{C}^{M}$  stands for the orthogonal projection  that  keeps the elements with index inside $\Omega$. Note that the energy of the noise content in $\tilde{\mathbf{y}}_{\Omega}$ is upperbounded by $\sqrt{|\Omega|}\eta = \sqrt{M}\, \eta$ with high probability.

The main challenge with signal model in \eqref{eq:measured_Discrete}, is to estimate $y(t)$, or equivalently, $\{b_k\}_{k\in [K]}$ and $\{z_k\}_{k\in [K]}$, by observing the noiseless samples $\mathbf{y}_{\Omega}$ or the noisy measurements $\tilde{\mathbf{y}}_{\Omega}$. In this work, our focus is on the case where $K \ll N$. 
This challenge appears in a number of real-world applications, such as
magnetic resonance imaging (MRI) and X-ray computed tomography 
\cite{jin2016general}, direction of arrival estimation \cite{Bokaei2022Hankel},  communication systems~\cite{razavikia2022blind}\cite{daei2023blind},  spike sorting in neural recordings \cite{spikesorting}, and super-resolution microscopy \cite{min2015fast}.
}

 
\section{Low Rank interpolation}\label{Sec:DOAestimation}

To recover the unseen samples from the available  measurements in subset  $\Omega$, one can use the fact that  the rank of the Hankel transform $\mathscr{H}(\mathbf{y})$ 
is upper-bounded by $K$ which is usually smaller than the size of $\mathbf{y}$ \cite{chen2014robust,razavikia2019reconstruction,razavikia2019sampling,ye2016compressive}.  Similar structures like Toeplitz, wrap-around Hankel, and double-Hankel impose similar low-rank properties. To include all these structures in our analysis, we use the generic $\Lif$ operator defined in \eqref{eq:DefLif} that maps 
 samples of exponential mixtures 
 into low-rank matrices. 
By choosing $\Lif$, our next step is to recover $\Lif(\mathbf{y})$ based on the measurements $\mathbf{y}_{\Omega}$ (or $\tilde{\mathbf{y}}_{\Omega}$). In other words, the measurements within the index set $\Omega$ shall be extended to the whole set $[N]$. As $\Lif(\mathbf{y})$ is a low-rank matrix, this task can be reformulated as a matrix completion problem: the elements of $\Lif(\mathbf{y})$ associated with $\Omega$ are observed (possibly noisy), and we want to estimate the rest. 

In a matrix completion problem, ideally, one searches for a matrix with the minimum rank that satisfies the constraints. The rank function is, however, both non-convex and non-smooth. Therefore, the exact rank minimization problem is generally NP-hard. 
The common alternative is to relax the ${\rm rank}(\cdot)$ function with the nuclear norm \cite{candes2009exact,recht2010guaranteed,gross2011recovering}. 
Adopting this relaxation, we shall consider the following matrix completion problem.
\begin{equation}
\label{convex-opt}
\begin{aligned}
\widehat{\mathbf{y}} = \underset{ \mathbf{g} \in \mathbb{C}^{N}}{\rm argmin} \quad \| \Lif({\mathbf{g}}) \|_{*}, \quad 
{\rm s.t.} \quad  \mathcal{P}_{\Omega}( \mathbf{g})  = \mathbf{y}_{\Omega},
\end{aligned}
\end{equation}
for the noiseless, and 
\begin{equation}
\label{convex-noisy-opt}
\begin{aligned}
\widehat{\tilde{\mathbf{y}}} = \underset{ \mathbf{g} \in \mathbb{C}^{N}}{\rm argmin} \quad  \| \Lif({\mathbf{g}}) \|_{*}, \quad 
{\rm s.t.} \quad \|\mathcal{P}_{\Omega}( \mathbf{g})- \tilde{\mathbf{y}}_{\Omega}\|_2 \leq \sqrt{M}\eta,
\end{aligned}
\end{equation}
for the noisy case, $\eta > 0$ was previously introduced as an upper bound for noise amplitudes (with high probability).

\subsection{Weighted Matrix Completion}

In our scenario, the samples are fixed, and we cannot sample adaptively. As a result, the conventional interpretation of  leverage scores as the sampling probabilities is useless here \cite{chen2015completing}.
{\changeB Instead, we try to set two weight matrices $\mathbf{W}_{L} \in \mathbb{C}^{d_1\times d_1}$ and $\mathbf{W}_{R} \in \mathbb{C}^{d_2\times d_2}$ such that the recovery of $\mathbf{W}_{L}\Lif({\mathbf{g}})\mathbf{W}_{R}^{\mathsf{H}}$ with a non-uniform sampling strategy that is consistent with the available samples (a more rigorous definition will be provided later) requires fewer samples.} The weighted lifted-structured low-rank matrix recovery is defined by incorporating left and right weight matrices into \eqref{convex-opt} as
\begin{equation}
\label{weighted-convex-opt}
\begin{aligned}
\widehat{\mathbf{y}} = \underset{ \mathbf{g} \in \mathbb{C}^{N}}{\rm argmin}&
& & \| \mathbf{W}_{L}\Lif({\mathbf{g}})\mathbf{W}_{R}^{\mathsf{H} }\|_{*} 
%
%
\quad  {\rm s.t.}~~~&
& &  \mathcal{P}_{\Omega}( \mathbf{g})  = \mathbf{y}_{\Omega}.\\  
\end{aligned}
\end{equation}
Similarly, for the noisy case, we have that
\begin{equation}
\label{weighted-convex-noisy}
\begin{aligned}
\widehat{\mathbf{y}} = \underset{ \mathbf{g} \in \mathbb{C}^{N}}{\rm argmin}&
& & \| \mathbf{W}_{L}\Lif({\mathbf{g}})\mathbf{W}_{R}^{\mathsf{H} }\|_{*} 
%
%
\quad  {\rm s.t.}~~~&
& & \| \mathcal{P}_{\Omega}( \mathbf{g} ) -\tilde{\mathbf{y}}_{\Omega} \|_{\rm F} \leq  \sqrt{M}\eta. 
\end{aligned}
\end{equation}
 {\changeB The optimization in \eqref{weighted-convex-noisy} is convex and can be reformulated into a semi-definite program (SDP) using Schur complement as in \cite{recht2010guaranteed}, i.e.,
\begin{equation}
\label{SDP-opt}
\begin{aligned}
\widehat{\mathbf{y}} = \underset{ \mathbf{P}, \mathbf{Q}}{\rm min}&
&& \tfrac{1}{2} {\rm tr}(\mathbf{P}) + \tfrac{1}{2} {\rm tr}(\mathbf{Q}) \\
{\rm s.t.}~~~&
&& \left[ \begin{matrix}
\mathbf{P} & \mathbf{W}_R \mathscr{L}^{\mathsf{H}}(\mathbf{g})\mathbf{W}_L^{\mathsf{T}} \\
\mathbf{W}_L \mathscr{L}(\mathbf{g})\mathbf{W}_R^{\mathsf{T}} & \mathbf{Q}
\end{matrix}
\right]  \succeq \mathbf{0}, \\
~~~& 
&& \mathcal{P}_{\Omega}(\mathbf{g}) = \mathbf{y}_{\Omega}, \quad \mathbf{P}, \mathbf{Q} \succeq \mathbf{0},
\end{aligned}
\end{equation}
where $\mathbf{P}\in \mathbb{C}^{d_2\times d_2}$ and $\mathbf{Q}\in \mathbb{C}^{d_1\times d_1}$ are Hermitian matrices.}

We should highlight that the results in \cite{chen2014robust} are not directly applicable to the weighted problem in \eqref{weighted-convex-opt}. In the next section, we analyze the weighted minimization with the general perspective of non-uniform sampling.

\section{Theoretical Guarantee and Main results}
\label{sec:Theoritical}

In this section, we investigate the conditions under which the uniqueness of the solution can be guaranteed. 
Here, we present recovery guarantees for generic lifting operators $\Lif$ that transform the vector of exponential mixtures  into low-rank matrices (including Hankel, double Hankel,  wrap-around Hankel, Toeplitz, Hankel-block-Hankel, and multi-level Toeplitz among others).

Before proceeding further, 
let us define the weighted leverage scores  as a generalization of 
Definition \ref{LEV_SCORS}.

\begin{definition}
\label{LEV_SCORS_M}
For given weight matrices $\mathbf{W}_{L}, \mathbf{W}_{R}$ and an arbitrary vector  $\mathbf{x} \in \mathbb{C}^{N}$,  assume  that the rank of $\mathbf{W}_{L} \mathscr{L}(\mathbf{x}) \mathbf{W}_{R}^{\mathsf{H}}$ is  $\tilde{K}$. Further, let ${\mathbf{U}}_{d_1\times \tilde{K}}{\boldsymbol{\Sigma}}_{\tilde{K}\times \tilde{K}} ({\mathbf{V}}_{d_2\times \tilde{K}})^{\mathsf{H}}$ be the SVD thereof. 
For each $n \in[N]$, we define the weighted leverage scores  $\tilde{\mu}_{n}$ as
\begin{align}
\label{eq:mutildDef}
\tilde{\mu}_{n} :=\frac{N}{\tilde{
K}} \max\{\|  \mathcal{P}_{U} (\mathbf{A}_{n})\|_{\rm F}^2,\| \mathcal{P}_{V}(\mathbf{A}_{n}) \|_{\rm F}^2   \},  \quad n \in[N],
\end{align}
where $\mathcal{P}_{U}(\mathbf{Y})$ and $\mathcal{P}_{V}(\mathbf{Y})$ for arbitrary $\mathbf{Y} \in \mathbb{C}^{d_1 \times d_2}$ are defines as:
\eas{
\mathcal{P}_{U}(\mathbf{Y}) &= \mathbf{W}_{L}^{\mathsf{H}} \mathbf{U} \left( \mathbf{U}^{\mathsf{H}} \mathbf{W}_{L} \mathbf{W}_{L}^{\mathsf{H} } \mathbf{U} \right)^{-1} \mathbf{U}^{\mathsf{H}} \mathbf{W}_{L}\mathbf{Y},\\ 
\mathcal{P}_{V}(\mathbf{Y}) &=  \mathbf{Y}\mathbf{W}_{R}^{\mathsf{H}} \mathbf{V} \left( \mathbf{V}^{\mathsf{H}} \mathbf{W}_{R} \mathbf{W}_{R}^{\mathsf{H} } \mathbf{V} \right)^{-1} \mathbf{V}^{\mathsf{H}} \mathbf{W}_{R}.
}{}
\end{definition}

\subsection{Recovery Guarantees for Non-uniform Random Sampling}\label{subsec:Gaurantee}
We first assume a random sampling scenario in which $\Omega$ is formed by selecting $n\in[N]$ with probability $p_n$ independently of other elements $k\neq n$. Below, we provide a set of lower bounds on $\{p_n\}$s to guarantee perfect (or robust) recovery with a high probability of using noiseless (noisy) samples.

\begin{theorem}\label{th:recovery}
{ \changeB Let $\mathbf{y}\in\mathbb{C}^{N}$ be as in \eqref{eq:measured_Discrete}, and $\Omega$ represent a location set of size $M$ formed by selecting each element $n\in[N]$  with probability $p_n$ independent of other elements.
We can recover $\mathbf{y}$ from the measurements   $\mathbf{y}_{\Omega} = \mathcal{P}_{\Omega}(\mathbf{y})$ using the noiseless setup in \eqref{weighted-convex-opt} with  probability no less than $1-N^{3-b_1}$ if}
\begin{align}\label{eq:sample_com_exact}
p_n \geq 
\min \lcb 1 \,,\, \f 1 N \max \lcb 1 \,,\,  \Rsf_{\mathscr{L}}^2 c\, {\tilde{\mu}}_{n} \Kt^2 \log(N) \rcb\rcb,
\end{align}
and 
\begin{align}
\label{eq:Cond2}
\frac{1}{8\sqrt{\log(N)}}\leq\min_{i\in [N]}\Big\{ \|\mathbf{A}_{i}\|_{0}\min\{\|\mathcal{P}_{U}(\mathbf{A}_{i})\|_{\rm F}^2, \|\mathcal{P}_{V}(\mathbf{A}_{i})\|_{\rm F}^2\} \Big\},
\end{align}
where $d_1$, $d_2$ and $\tilde{K}$ are dimensions and the rank of the lifted structure $\mathscr{L}(\mathbf{y})$, respectively. 
Additionally, $\tilde{\mu}_{n}$ is the weighted leverage score in \eqref{eq:mutildDef}, the coefficient $\Rsf_{\mathscr{L}}$ is defined as 
\begin{align}
\label{eq:basis_matrices_property}
    \Rsf_{\mathscr{L}}
    =  \sum_{n \in [N]}  \big\|\mathbf{A}_{n} \odot
    \mathbf{A}_{n}\big\|_{\infty \rightarrow \infty},
\end{align}
and $c =  192^2(b_1+1)$ for $b_1\geq 3$.
\end{theorem} 

\begin{proof}
The proof is provided in the appendix.
\end{proof}

\begin{corollary}
The parameter $\Rsf_{\mathscr{L}}$ in \eqref{eq:basis_matrices_property} in Theorem  \ref{th:recovery} is a function of the  lifted-structure. 
For instance, $\Rsf_{\mathscr{L}} = \Ocal(\log{(N)})$ for  Hankel, Toeplitz, and double-Hankel structures. For the wrap-around Hankel structure, however, we have $\Rsf_{\mathscr{L}} = \Ocal(1)$.
\end{corollary}

\begin{remark}
\label{remark1}
{\changeB The expected number of observed elements $|\Omega|$ in Theorem  \ref{th:recovery} is no less than $c {\Rsf_{\mathscr{L}}^2} \big(\frac{\sum_{n}\tilde{\mu}_{n}}{N}\big)\tilde{K}^2\log{(N)}$. With Hoeffding’s inequality, it is possible to check that the actual number of observed elements in this random setting  concentrates around its expected value and is upper bounded by $M \leq 2 c {\Rsf_{\mathscr{L}}^2} \big(\frac{\sum_{n}\tilde{\mu}_{n}}{N}\big)\tilde{K}^2\log{(N)}.$ with high probability. } 
\end{remark}

\begin{figure}[t!]
\centering
\begin{subfigure}{.5\linewidth}
\centering
\begin{tikzpicture}
\begin{axis}[
xlabel={Number of Samples},
ylabel={Sparsity level},
title = {WLi-DEMaC},
width=7 cm,
height=6 cm,
legend pos= south east,
legend style={{font=\tiny} },
enlargelimits=false,
colormap/blackwhite]
    \addplot [
    matrix plot*,
    mesh/cols=19,
    point meta=explicit, forget plot
    ] table [meta=C] {figs/phase_trans/WDEMaC.dat};
    \addplot[samples=10, domain=2:53, name path=A,color=brown,style=dotted, ultra thick]{x^1.5/20+1)};
    \addplot[samples=10, domain=2:56, name path=A,color=blue,style=dashed, ultra thick] {x^1.5/21+0.5)};
         \legend{            
            WLi-DEMaC,
            DEMaC,
        }
\end{axis}
\end{tikzpicture}
\centerline{{(a)}} \medskip
\end{subfigure}
\begin{subfigure}{.49\linewidth}
\centering
\begin{tikzpicture}
\begin{axis}[
xlabel={Number of Samples},
title = {WLi-EMaC},
width=7 cm,
height=6 cm,
legend pos= south east,
legend style={{font=\tiny} },
enlargelimits=false,
colorbar,
colorbar style={
     width=0.05*\pgfkeysvalueof{/pgfplots/parent axis width},
},
colormap/blackwhite]
    \addplot [
    matrix plot*,
    mesh/cols=19,
    point meta=explicit, forget plot, 
    ] table [meta=C] {figs/phase_trans/WEMaC.dat};
\addplot[samples=10, domain=2:57, name path=A,color=lime,style=dotted,ultra thick] {x^2/170+1)};
\addplot[samples=10, domain=2:58, name path=A,color=red,style=dashed, ultra thick] {x^2/200+0.5)};
         \legend{            
            WLi-EMaC,
            EMaC,
        }
\end{axis}
\end{tikzpicture}
\centerline{{(b)}} \medskip
\end{subfigure}
\begin{subfigure}{.5 \linewidth}
\centering
\begin{tikzpicture}
\begin{axis}[
xlabel={Number of Samples},
ylabel={Sparsity level},
title = {DEMaC},
width=7 cm,
height=6 cm,
enlargelimits=false,
colormap/blackwhite]
    \addplot [
    matrix plot*,
    mesh/cols=19,
    point meta=explicit,
    ] table [meta=C] {figs/phase_trans/DEMaC.dat};
\addplot[samples=10, domain=2:56, name path=A,color=blue,style=dashed, ultra thick] {x^1.5/21+0.5)};
 \addplot[samples=10, domain=2:53, name path=A,color=brown,style=dotted, ultra thick]{x^1.5/20+1)};
\end{axis}
\end{tikzpicture}
\centerline{{(c)}} \medskip
\end{subfigure}
\begin{subfigure}{.49\linewidth}
\centering
\begin{tikzpicture}
\begin{axis}[
xlabel={Number of Samples},
title = {EMaC},
width=7 cm,
height=6 cm,
enlargelimits=false,
colorbar,
colorbar style={
     width=0.05*\pgfkeysvalueof{/pgfplots/parent axis width},
},
colormap/blackwhite]
    \addplot [
    matrix plot*,
    mesh/cols=19,
    point meta=explicit,
    ] table [meta=C] {figs/phase_trans/EMaC.dat};
\addplot[samples=10, domain=2:57, name path=A,color=lime,style=dotted,ultra thick] {x^2/170+1)};
\addplot[samples=10, domain=2:58, name path=A,color=red,style=dashed, ultra thick] {x^2/200+0.5)};
\end{axis}
\end{tikzpicture}
\centerline{{(d)}} \medskip
\end{subfigure}

\vspace{-0.75cm}
\caption{(a) WLi-DEMaC -- (b) WLi-EMaC -- (c) DEMaC -- (d) EMaC phase transition diagrams in Section \ref{Sec:Simulation} as a function of the  number of samples -- $x$ axis--,  $M$ in \eqref{eq:Projection}, and the sparsity level -- $y$ axis--, evaluated as number of frequency components ($K$) in  \eqref{eq:measured_Discrete}. In (a) and (c), the dotted brown and the dashed blue lines approximately show the boundary of the transition between the success and failure cases for WLi-DEMaC and DEMaC, respectively. Similarly, in (b) and (d) dotted green and  dashed red lines show the transition boundary between the success and failure cases for WLi-EMaC and EMaC, respectively.
}
\label{fig:phaseTrans}
\end{figure}
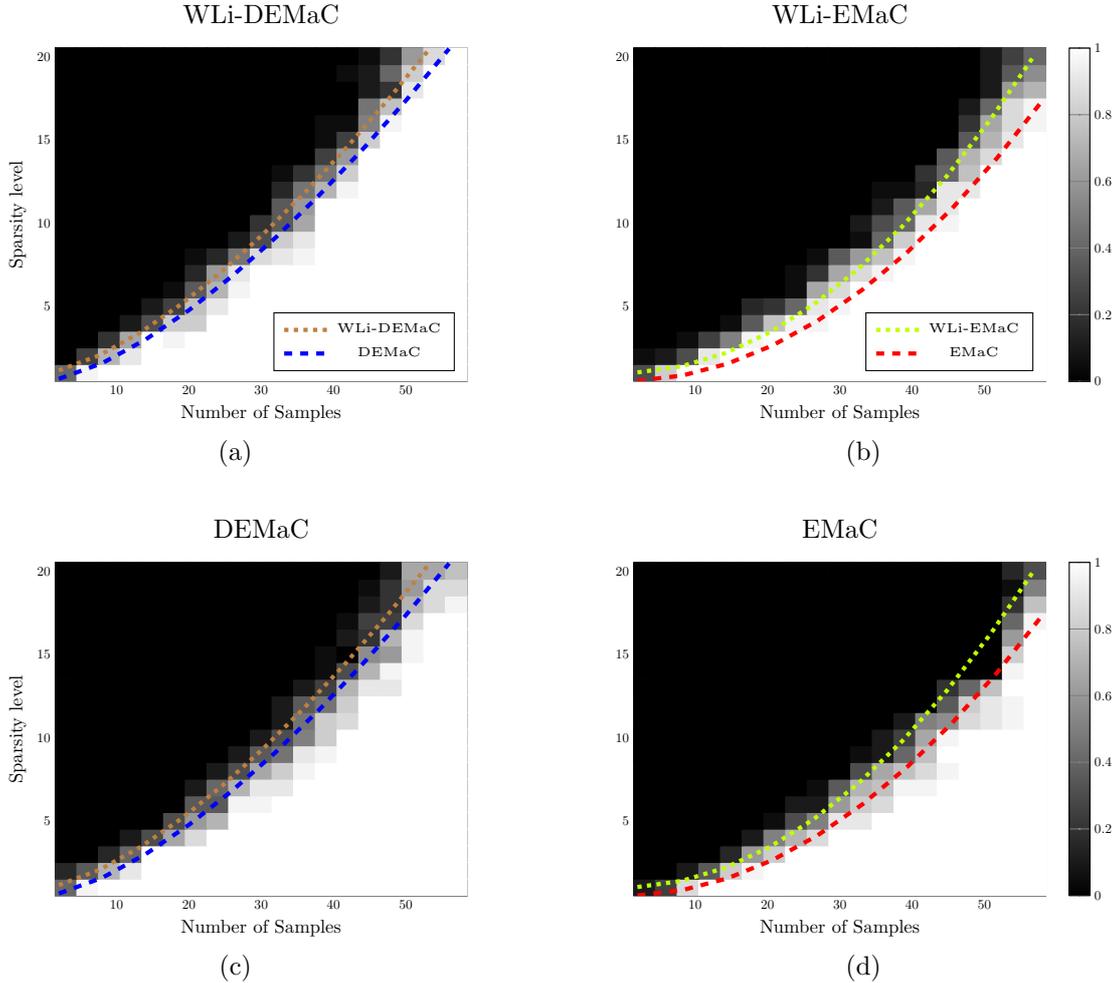

\begin{remark}
\label{remark2}
If $\mathscr{L}$ is the Hankel structure, we call the resulting method WLi-EMaC (because of the similarity of the technique with EMaC in \cite{chen2014robust}).
The guaranteed sample size for exact recovery using WLi-EMaC and EMaC algorithms are $\Rsf_{\mathscr{L}}^2 c\big(\frac{\sum_{n}\tilde{\mu}_{n}}{N}\big)\tilde{K}^2\log{(N)}$ and $c_1 \max\{\mu_n\}\tilde{K}\log^4{(N)}$, respectively. 
As for the contribution of the leverage scores, we should emphasize that $\max\{\mu_n\}$ in EMaC is reduced to $ \frac{\sum_{n}\tilde{\mu}_{n}}{N} $ in WLi-EMaC (the average score instead of the maximum score). 
This reduction is substantial  when some complex basis of the input signal are similar to each other, i.e., $z_k\approx z_{k'}$ for $k,k' \in [K]$.
In such cases, few leverage scores become very large, causing a considerable gap between the maximum and the average leverage scores. Although the guaranteed sample size scales with $\tilde{K}^2$ in WLi-EMaC (compared to $\tilde{K}$ in EMaC), our numerical results in the form of the phase transition diagrams in Figure \ref{fig:phaseTrans} show that the actual sample size scales with $K$ and not $\tilde{K}^2$. We believe there is room for improvement in our theoretical analysis of the guarantees.

\end{remark}
 
{\changeB
\begin{remark}\label{remark3}
    The constraint in \eqref{eq:Cond2} captures the incoherence condition in low-rank recovery problems that reflects both the characteristics of the lifting structure -- basis $\bm{A}_i$ --  and the location of the frequencies -- $\mathcal{P}_{\bm{U}}$ and $\mathcal{P}_{\bm{V}}$.
\end{remark}
}
In Theorem \ref{th:noisy_recovery}, a linear error bound for the recovery in terms of the input noise level is established.
\begin{theorem}
\label{th:noisy_recovery}
Let $\mathbf{y} $ and $\Omega$ be similarly defined to Theorem \ref{th:recovery} and  assume \eqref{eq:sample_com_exact} holds. Further, let $\tilde{\mathbf{y}}_{\Omega} = P_{\Omega}( \tilde{\mathbf{y}} )$ be the vector of observed noisy measurements where the noise term $ \mathbf{e} \in \mathbb{C}^{N} $  satisfies $\| P_{\Omega}( \mathbf{e} )\|_{2} \leq \sqrt{M} \eta$. Then, with a probability no less than $ 1-N^{3-b_1} $ ($ b_1\geq 3$), any solution $ \widehat{\mathbf{y}} $ of  \eqref{weighted-convex-noisy} satisfies
		\begin{equation}
\Big\| \mathbf{W}_L\big(\mathscr{L}{(\widehat{\mathbf{y}})} - \mathscr{L}{(\mathbf{y})}\big)\mathbf{W}_R^{ \mathsf{H} } \Big\|_{\rm F} \leq  c_2 \sqrt{M}\eta \tfrac{\min(d_1, d_2)}{\min_{n} p_n^2},
\end{equation}
where  $c_2<102$ is a fixed constant. 
\end{theorem}
\begin{proof}
The proof is provided in the supplementary material, Sec. II.
\end{proof}

\begin{corollary}
\label{col:diag_weight}
In Definition \ref{LEV_SCORS_M}, if $\mathbf{W}_L \in \mathbb{R}_+^{d_1\times d_1}$ and $\mathbf{W}_R \in \mathbb{R}_+^{d_2\times d_2}$ are restricted to  non-negative-valued diagonal matrices,  i.e.,
\begin{align}
\mathbf{W}_L &= {\rm diag}\big(\sqrt{w_{L,1}},\dots \sqrt{w_{L,d_1}}\big), ~~~
\mathbf{W}_R = {\rm diag}\big( \sqrt{w_{R,1}} ,\dots, \sqrt{w_{R,d_2}}\big),
\end{align}
then, the leverage scores will be bounded by
\begin{align}
\label{diag_modified_lev_scores}
\frac{\tilde{\mu}_{n}\tilde{K}}{N} \leq  \max \Bigg\{\frac{\left\| \mathbf{W}_{L} \mathbf{A}_n \right\|_{\rm F}^2 }{\sum_{k=1}^{ \lfloor \frac{N}{\beta \tilde{K}} \rfloor}   w_{L,i_k} } , \frac{\left\| \mathbf{A}_n \mathbf{W}_{R}^{\mathsf{T} }  \right\|_{\rm F}^2 }{ 
\sum_{k=1}^{ \lfloor \frac{N}{\beta \tilde{K}} \rfloor  } w_{R,j_k} }\Bigg\},
\end{align}
where  {$w_{L,i_1}  \leq \ldots \leq w_{L,{i_{d_1}}} $ and $w_{R,j_{1}}  \leq \ldots \leq w_{R,j_{d_2}}$} are the sorted squared diagonal elements of $\mathbf{W}_{L}$ and $\mathbf{W}_{R}$, respectively, and $ \beta = \tfrac{N}{\tilde{K}}\max \lcb \frac{1}{\| \mathbf{U}^{\mathsf{H}} \|^2},  \frac{1}{\| \mathbf{V}^{\mathsf{H}} \|^2} \rcb$. 
\end{corollary}
\begin{proof}
The proof is provided in the supplementary material, Sec. VII.
\end{proof}

{ \changeA 

\subsection{Adjusting the Weight Matrices}\label{subsec:WeightTune}

On the one hand, Theorems \ref{th:recovery}  and \ref{th:noisy_recovery} reveal a linear relationship between the guaranteed sample complexity for perfect recovery and the leverage scores.
On the other hand, the leverage scores are upper-bounded in Corollary \ref{col:diag_weight}  by the weight matrices. Therefore, one can adjust the weight matrices such that the overall leverage scores are minimized; this, in turn, reduces the upper bound on the sample complexity. 
In the deterministic setup, the strategy is to interpret the actual $\Omega$ as a realization of a random sampling set with element-wise probabilities $\{p_n\}_{n \in[N]}$. If the probabilistic guarantee works for $\{p_n\}_{n \in[N]}$, then $\Omega$ as a realization of that random sampling is also suitable with high probability. 

 Next, we maximize the likelihood of the observed samples by tuning the set $\{p_n\}_{n \in [N]}$. Our approach is to determine weight matrices $\mathbf{W}_L,\mathbf{W}_R$ such that the likelihood of observing $\Omega$ attains its maximum point in one of the suitable random sampling strategies given in \eqref{eq:sample_com_exact}.}
 This strategy results in
\begin{align}
\label{weight_cal_1}
{\mathbf{W}}_{L}, {\mathbf{W}}_{R}   = &\underset{\substack{ \mathbf{W}_{L} \in \mathbb{C}^{d_1\times d_1},\\ \mathbf{W}_{R} \in  \mathbb{C}^{d_2 \times d_2}} }{\rm argmax} 
-\sum_{n \not\in \Omega} p_{n} 
\equiv \underset{\substack{ \mathbf{W}_{L} \in \mathbb{C}^{d_1\times d_1},\\ \mathbf{W}_{R} \in  \mathbb{C}^{d_2 \times d_2}} }{\rm argmin} 
\sum_{n \not\in \Omega} \tilde{\mu}_{n}.
\end{align}
By solving the aforementioned optimization problem, we obtain weight matrices that reduce the sample complicity or, alternatively, increase the reconstruction quality. 
We should note that in cases where prior knowledge about the subspace of the signal is available (i.e., $\mathbf{U}$ and $\mathbf{V}$ are known), one can set the weight matrices to maximize the reconstruction quality  \cite{ardakani2022multi}.  In this work, however, such information is not available. {\changeB Instead, in \cite[Section 4]{partII}, we devise an optimization problem  to solve the maximization in \eqref{weight_cal_1} and describe the whole procedure in detail.}


\section{Numerical Simulations}
\label{Sec:Simulation}

We consider two options for the lifting operator $\mathscr{L}$ in this work:
Hankel and double Hankel. We refer to these two implementations of the proposed algorithm as  WLi-EMaC and WLi-DEMaC  because of their similarity to EMaC and DEMaC in \cite{chen2014robust} and \cite{yang2021new}, respectively.
In the sequel, we present numerical experiments  comparing WLi-EMaC and WLi-DEMaC against their non-weighted counterparts EMaC \cite{chen2014robust} and DEMaC \cite{yang2021new}.
{\changeB We set $N = 59$ to be an odd integer which enables us to set the Hankel pencil parameter $d_1 = d_2 = 30$ so as to achieve a square matrix. For the same reason, in the DEMaC structure, we set $d_1 = d_2 = 40$.}




In Figure \ref{fig:phaseTrans}, we plot the  phase transition diagram for WLi-EMaC, WLi-DEMaC, DEMaC, and EMaC algorithms.
For these numerical evaluations, all algorithms are implemented by CVX toolbox with SDPT3 solver in MATLAB \cite{cvx}. 
For each pair of $(M, K)$, the success rates are averaged over $100$ Monte Carlo trial iterations. A  trial is  considered successful  if it satisfies $\frac{\| \mathbf{x} - \mathbf{\hat{x}} \|_{\rm F}}{\| \mathbf{x} \|_{\rm F}} \leq 10^{-3} $, where $\mathbf{x}$ and $\mathbf{\hat{x}}$ are 
the ground truth signal and its estimated vector, respectively.
The average success rates  for each cell  are depicted 
in Figure \ref{fig:phaseTrans}: brighter colors mean higher average success rates.

Results in Figure \ref{fig:phaseTrans} imply that the required sample size $M$ for exact recovery with both the Hankel and double Hankel structures is proportional to $K$ in the noiseless setting.  The dashed lines almost depict the transition boundary between the success and failure cases for weighted versus unweighted strategies.

This reveals that our algorithm performs better than the predicted  bound in Theorem \ref{th:recovery}, where the required sample size $M$ for exact recovery scales with $K^2$.
{\change 
By comparing the phase transitions in Figure \ref{fig:phaseTrans}, we observe that the proposed weighted approach improves the completion performance  for both Hankel and Double-Hankel structures. Indeed, top row charts in Figure \ref{fig:phaseTrans} show the superiority of the WLi-DEMaC and the WLi-EMaC algorithms over the DEMaC and the EMaC algorithms, respectively. Also, Figure \ref{fig:phaseTrans} indicates that the Double Hankel structure results in better reconstructions compared to the simple Hankel structure.
}

\section{Conclusion}
\label{Sec:Conclude}
{\change 
In this paper we proposed a novel approach for recovering the summation of exponential functions closely related to the line spectral estimation problem. The proposed approach comprised of three steps: 1) lifting the observed samples to a chosen structured matrix such as Hankel or Toeplitz, 2) tuning the left and right weighting matrices based on the sample informativeness, and 3) solving a weighted matrix completion problem to find the missing samples. 
%
For a given choice of the lifting structured matrix, this weighting approach generalizes  other low-rank matrix completion techniques  in the literature, such as EMaC (Hankel) and DEMaC (double Hankel). Both theoretical analysis and numerical results showed that the weighted lifted (WLi-) approach outperforms the case without weighting. In other words, WLi-EMaC and WLi-DEMac outperform EMaC and DEMaC in terms of NMSE.
}

\appendix
\section{Proof of Theorem \ref{th:recovery}}
In this appendix, we shall present the proof of Theorem  \ref{th:recovery}. 
We do so by analyzing the dual problem.

More specifically, we will construct an appropriate dual certificate; the existence of this certificate guarantees that the solution to the problem. To prove the uniqueness of the solution, we use the well-studied golfing scheme, first used in \cite{gross2011recovering} to verify the existence of an approximate dual certificate. 
As the first step, we define the sampling operator $ \mathcal{A}_{n}$ for any matrix $\mathbf{M} \in \mathbb{C}^{d_1\times d_2}$ as follows:
\begin{equation}
\mathcal{A}_{n}(\mathbf{M})=  \big\langle\mathbf{M},\mathbf{A}_{n}\big\rangle \mathbf{A}_{n} = {\rm tr}\big(\mathbf{M}^{\mathsf{T}}\mathbf{A}_{n}\big) \mathbf{A}_{n}.
\end{equation}

Let $\Omega$ be a random subset of $[N]$ such that the element $1\leq n\leq N$ appears in $\Omega$ with probability $p_n$ independent of other elements. We define the projection operator onto $\Omega$ as
\begin{align}
\mathcal{A}_{\Omega}=& \sum_{n \in [N]} \frac{\delta_{n}}{p_{n}} \mathcal{A}_{n}.
\end{align}
where $\delta_{n}$ is  equal to $1$ for $n \in \Omega$ and zero elsewhere and $ p_{n} $ is sampling probability of $n$-th element. We can check that $\mathbb{E}[\mathcal{A}_{\Omega}] = \mathcal{A}$, where $\mathcal{A}$ stands for $\sum_{n=1}^{N}\mathcal{A}_{n}$. It is also simple to verify that
\begin{align}
\|\mathcal{A}_{\Omega}\| & = \Big\|\sum_{n \in [N]} \frac{\delta_{n}}{p_{n}} \mathcal{A}_{n} \Big\| 
\leq  \tfrac{1}{\min_n  \{ p_n\}}.
\label{eq:AOmega_norm}
\end{align}

The projection definition of $\mathcal{A}_{\Omega}$ implies that for all $\Omega$, the operator $\mathcal{A}_{\Omega}$ is a self-adjoint operator. Now, we reformulate the main problem in \eqref{weighted-convex-opt} as a matrix recovery problem in the lifted domain  ($\mathscr{L}$). We define $ \mathbf{U} $ and $ \mathbf{V}$ as the left and right unitary matrices in the reduced SVD of  $\mathbf{W}_{L}  \mathscr{L}(\mathbf{M}) \mathbf{W}_{R}^{\mathsf{H}} = \mathbf{U} \boldsymbol{\Sigma}  \mathbf{V}^{\mathsf{H}}$. Moreover, for every $\mathbf{Y} \in \mathbb{C}^{ d_1 \times d_2}$, $\mathcal{P}_{U}$ and $\mathcal{P}_{V}$ are defined as $\mathcal{P}_{U}(\mathbf{Y}) = \mathbf{W}_{L}^{\mathsf{H} } \mathbf{U} \left( \mathbf{{U}}^{\mathsf{H}} \mathbf{W}_{L} \mathbf{W}_{L}^{\mathsf{H} } \mathbf{{U}} \right)^{-1} \mathbf{{U}}^{\mathsf{H}} \mathbf{W}_{L}  \mathbf{Y} $ and $ \mathcal{P}_{V}(\mathbf{Y}) = \mathbf{Y} \mathbf{W}_{R}^{\mathsf{H} } \mathbf{V} \left( \mathbf{V}^{\mathsf{H}} \mathbf{W}_{R} \mathbf{W}_{R}^{\mathsf{H} } \mathbf{V} \right)^{-1} \mathbf{V}^{\mathsf{H}} \mathbf{W}_{R} $ respectively. A simple matrix multiplication shows that for all $\mathbf{Y}$, we have
\begin{align}\label{eq:PU}
\mathbf{{U}}^{\mathsf{H}} \mathbf{W}_{L} \mathcal{P}_{U}(\mathbf{Y}) &= \mathbf{{U}}^{\mathsf{H}} \mathbf{W}_{L} \mathbf{Y},\\
\mathcal{P}_{V}(\mathbf{Y}) \mathbf{W}_{R}^{\mathsf{H} } \mathbf{V}  &= \mathbf{Y} \mathbf{W}_{R}^{\mathsf{H} } \mathbf{V} . \label{eq:PV}
\end{align}
We define the orthogonal operator as $\mathcal{A}^{\perp} = \mathcal{I} -  \mathcal{A}$ where $\mathcal{I}$ is the identity operator. Then the tangent space $T$ with respect to  $ \mathbf{W}_{L} \mathscr{L}(\mathbf{M}) \mathbf{W}_{R}^{\mathsf{H}} = \mathbf{U} \boldsymbol{\Sigma} \mathbf{V}^{\mathsf{H}} $ is defined as
\begin{equation}
\begin{aligned}
T := \{ & {\mathbf{W}_{L}^{\mathsf{H} } \mathbf{U}}\mathbf{Y}_{1}^{\mathsf{H}} + \mathbf{Y}_{2}\mathbf{V}^{\mathsf{H}} \mathbf{W}_{R} 
:\mathbf{Y}_1 \in  \mathbb{C}^{d_1\times \tilde{K}}, \mathbf{Y}_2 \in  \mathbb{C}^{d_2\times \tilde{K}}   \}.
\end{aligned}
\end{equation}
Also, the projection of a matrix $\mathbf{Z} \in \mathbb{C}^{d_1\times d_2} $ onto the tangent space is denoted by $\mathcal{P}_{T}(\mathbf{Z})$ and we have: 
\begin{align}\label{eq:PTdef}
\mathcal{P}_T(\mathbf{Z}) = \mathcal{P}_U(\mathbf{Z}) + \mathcal{P}_V(\mathbf{Z}) - \mathcal{P}_U(\mathcal{P}_V(\mathbf{Z})). 
\end{align}
We can now rewrite weighted lifted-structured low-rank matrix recovery problem in \eqref{weighted-convex-opt} in form of the following general matrix completion problem:
\begin{equation}
\label{eq:minimiztion}
\begin{aligned}
\widehat{\mathbf{M}} = \underset{ \mathbf{M} \in \mathbb{C}^{d_1\times d_2}}{\rm argmin}&
& & {\| \mathbf{W}_{L} \mathbf{M} \mathbf{W}_{R}^{\mathsf{H} } \|_{\rm *}} 
%
%
\quad
{\rm s.t.}~~~&
& & \mathcal{Q}_{\Omega}(\mathbf{M}) = \mathcal{Q}_{\Omega}( \mathscr{L}(\mathbf{y}) ),
\end{aligned}
\end{equation}
where $\mathcal{Q}_{\Omega}$ is defined as 
$\mathcal{Q}_{\Omega}  = \mathcal{A}_{\Omega} +  \mathcal{A}^{\perp}. $
Using \eqref{eq:AOmega_norm}, we can bound $\|\mathcal{Q}_{\Omega} \|$ as $\|\mathcal{Q}_{\Omega} \| \leq \|\mathcal{A}_{\Omega}\| + \| \mathcal{A}^{\perp} \| \leq \tfrac{1}{\min_n p_n} + 1$.

We further have $\mathbb{E}[\mathcal{Q}_{\Omega}] = \mathbb{E}[\mathcal{A}_{\Omega}] +  \mathcal{A}^{\perp} = \mathcal{A} +  \mathcal{A}^{\perp} = \mathcal{I}$.  As it can be seen in \eqref{eq:minimiztion}, scaling weight matrices does not change the problem's solution, and the matrices only need to be normalized. Hence, for simplicity of the proof, we assume  $\|\mathbf{W}_L\|_{\rm F} =\|\mathbf{W}_R\|_{\rm F} =  1.$

To prove the exact recovery of the convex optimization, it suffices to produce an appropriate dual certificate, as stated in the following lemma.
\begin{lemma} 
\label{main_lemma}
For a given $\Omega$, let the sampling operator $ \mathcal{Q}_{\Omega} $ fulfills
\begin{align}
\label{spectral_constraint}
{\| \mathcal{P}_{T} - \mathcal{P}_{T} \mathcal{Q} _{\Omega}\mathcal{P}_{T} \| \leq \frac{1}{2}} ,
\end{align} 
if there exists a matrix $\mathbf{G}$ satisfying
\begin{align}
\label{Lemma1_1}
\mathcal{Q}_{\Omega}^{\perp} (\mathbf{G}) = 0,
\end{align}
\begin{align}
\label{Lemma1_2}
{ \| \mathcal{P}_{T}(\mathbf{G}-\mathbf{W}_{L}^{\mathsf{H} }\mathbf{UV}^{\mathsf{H}} \mathbf{W}_{R}) \|_{\rm F} \leq  \frac{1}{5 \|\mathcal{Q}_{\Omega}\|} }  ,
\end{align}
and
\begin{align}
\label{Lemma1_3}
{ \| \mathcal{P}_{T^{\perp}}(\mathbf{G}) \| \leq \frac{1}{2}   },
\end{align}
 then, $\mathbf{M}$ is the unique solution to \eqref{eq:minimiztion}.
\end{lemma}
\begin{proof}
The proof is provided in the supplementary Section III.  
\end{proof}

Lemma \ref{main_lemma} will be satisfied, when it is  sufficiently incoherent respect to the tangent space $T$. we bound the fluctuation of $ \mathcal{P}_{T} \mathcal{A}_{\Omega} \mathcal{P}_{T} $ in the following lemma.

\begin{lemma}
	\label{lemma3}
	For a constant $0 < \epsilon \leq \frac{1}{2}$, if $ p_{n} \geq c_0 \frac{{\mu}_{n}r \log(N) } {N} $ for each $n \in [N]$ we have
	\begin{align}
	\left\| \mathcal{P}_{T} - \mathcal{P}_{T} \mathcal{Q} _{\Omega}\mathcal{P}_{T}\right\| \leq \epsilon
	\end{align}
	with probability exceeding $ 1 - N^{- b_1} $ for sufficiently large $ c_0 \geq \frac{56}{3} (b_1 + 1) $.	
\end{lemma}
\begin{proof}
	See supplementary Section IV.
\end{proof}

In what follows, we show there exist a dual certificate $\mathbf{G}$ such that it satisfies conditions in \eqref{Lemma1_2} to \eqref{Lemma1_3} with high probability. 

\subsection{Dual Certificates Construction} \label{sec:const_dual_cert}
We construct the dual certificate by using the golfing scheme introduced in \cite{gross2011recovering}. 
Let $ \epsilon < \frac{1}{e} $ be a small constant, and define $ L :=  \log_{\frac{1}{\epsilon}}({N^2}\| \mathcal{ {Q}}_{\Omega} \|) $.  
Let us form $L$ independent subsets $\{\Omega_{\ell}\}_{\ell=1}^{L}$ of $[N]$ by choosing the elements $1\leq n\leq N$ with probability $q_{n} := 1 - ( 1 - p_{n} )^{\frac{1}{L}}$ independent of each other. Further, let $\overline{\Omega} = \Omega_1 \cup \dots \cup \Omega_L$. We first check the probability that a given $1\leq n\leq N$ belongs to $\overline{\Omega}$:
\begin{align}
 \mathbb{P}[ n \in \overline{\Omega} ]  =1 - \prod_{\ell\in [L]} ( 1 - p_{n} )^{\frac{1}{L}} = p_n.
\end{align}
Hence, $\overline{\Omega}$ fulfils the required element-wise probabilities. 
Next, we construct the dual certificate matrix $ \mathbf{G} $  as 
\begin{align}\label{eq:Gdef}
\mathbf{G} := \sum_{\ell\in [L]} {\mathcal{Q}_{\Omega_{\ell}} (\mathbf{F}_{\ell})},
\end{align}
where $\mathbf{F}_{\ell} = \mathcal{P}_{T} (\mathcal{I}-\mathcal{Q}_{\Omega_{\ell}} ) \mathcal{P}_T(\mathbf{F}_{\ell-1})$ and  $\mathbf{F}_{0} = \mathbf{W}_{L}^{\mathsf{H} } \mathbf{UV}^{\mathsf{H}} \mathbf{W}_{R}$. 
Since $\mathbf{F}_{\ell} \in \overline{\Omega}$, we conclude that $\mathcal{Q}_{\overline{\Omega}}^{\perp} (\mathbf{G}) = 0$; i.e., $\mathbf{G}$ satisfies the first condition of Lemma \ref{main_lemma} for $\overline{\Omega}$.
In addition, we have that $
\mathcal{P}_{T} (\mathbf{F}_{\ell}) = \mathbf{F}_{\ell} = \Big( \mathcal{P}_{T} - \mathcal{P}_{T}\mathcal{Q}_{\Omega_{\ell}}  \mathcal{P}_T\Big)(\mathbf{F}_{\ell-1}).
$
Besides, from \eqref{lemma3}, we know that
\begin{align}
\Big\| \mathcal{P}_T - \mathcal{P}_T \mathcal{Q}_{\Omega_{\ell}}\mathcal{P}_T    \Big\|  \leq \epsilon < \frac{1}{2},
\end{align}
with a probability no less than $1-N^{-b_1}$.
To bound $ \big\| \mathcal{P}_T \big(\mathbf{G} - \mathbf{F}_{0} \big)  \big\|_{\rm F} $, we use a similar technique as in \cite{chen2014robust}  to obtain $\mathcal{P}_T (\mathbf{G} - \mathbf{F}_{0}) = - \mathcal{P}_T (\mathbf{F}_{L})$. The latter holds due to $ q_{\ell} \geq \tfrac{p_{\ell}}{L}\geq c_0 \mathsf{R}_{\mathscr{L}}^2  \frac{\tilde{\mu}_{\ell}\tilde{K}^2  } {N} $. Now, we are able to write
\begin{align}
\| \mathcal{P}_T \big(\mathbf{G} &- \mathbf{F}_{0} \big)  \|_{\rm F} =  \| \mathcal{P}_T \big(\mathbf{F}_{L} \big)  \|_{\rm F} \leq {\epsilon^L} \| \mathcal{P}_T(\mathbf{F}_0) \|_{\rm F}
<  \frac{1}{5\|\mathcal{Q}_{\Omega} \|},
\end{align}

with a probability no less than $1-L N^{-b_1}$. 
This shows that $\mathbf{G}$ satisfies condition \eqref{Lemma1_2} of Lemma \ref{main_lemma} with high probability. 


\subsection{Some relevant lemmas}
\label{sec:Some relevant lemmas}

We begin by defining the following two norms for the matrix $ \mathbf{M} \in \mathbb{C}^{d_1 \times d_2}$ for a given set of lifting basis   $\{\mathbf{A}_{n}\}_{n \in [N]}$: 

\eas{
\| \mathbf{M} \| _{\mathcal{A},\infty} 
& := \max_{n\in [N]} {\bigg|  \frac{{N} \langle\mathbf{A}_{n},\mathbf{M} \rangle }{\tilde{K} \tilde{\mu}_{n} \sqrt{\omega_{n}}  } \bigg|}  
\label{def_norm_max}, \\
 \| \mathbf{M} \| _{\mathcal{A},2} 
& :=  \sqrt{ \sum_{n\in [N]} {\frac{| {N} \langle \mathbf{A}_{n}, \mathbf{M} \rangle |^2  }{\tilde{K} \tilde{\mu}_{n} \omega_{n} }} },
\label{def_norm_2} 
}{\label{eq:norms}}
where we have defined   $\omega_n:=\|\mathbf{A}_n\|_0$.

We now state three inequalities regarding the  norms in \eqref{eq:norms}: Lemma  \ref{lemma4}, Lemma \ref{lemma5}, and Lemma \ref{lemma_6}.
Generally, these proofs rely on  matrix concentration inequalities in \cite[Appendix~A]{chen2014robust}.

\begin{lemma}
\label{lemma4}
Suppose $ \mathbf{M}$ is a complex-valued ${d_1 \times d_2}$ matrix. If  $ p_{n} \geq c_0 \frac{\tilde{\mu}_{n} \tilde{K}^2 \log(N)}{N} $ for all $n \in [N]$, then
\ea{
& \big\| \big({\mathcal{Q}_{\Omega} -  \mathcal{I} }  \big) \mathbf{M} \big\|  
\leq  \sqrt{\tfrac{2(b_2+1)}{c_0  \tilde{K} \mathsf{R}_{\mathscr{L}}}}  \| \mathbf{M} \|_{\rm \mathcal{A},2}  + \tfrac{2(b_2+1)}{3c_0\tilde{K} \mathsf{R}_{\mathscr{L}}} \| \mathbf{M} \|_{\rm \mathcal{A},\infty}, 
\nonumber
}

holds with a probability at least $1-N^{-b_2}$, where $ c_0 \geq 2(b_2 + 1)$ and ${\mathsf{R}}_{\mathscr{L}} = \sum_{n \in [N]}  \big\|\mathbf{A}_{n} \odot
    \mathbf{A}_{n}\big\|_{\infty \rightarrow \infty}$.
\end{lemma}
\begin{proof}
See supplementary Section V. 
\end{proof}
We further control $ \|.\|_{\mathcal{A}, 2} $ and $ \|.\|_{\mathcal{A}, \infty} $ norms of $ \big( {\mathcal{P}_T \mathcal{Q}_{\Omega}} - \mathcal{P}_T  \big) $ in the next two lemmas.

\begin{lemma}
\label{lemma5}
For $ c_0 \geq 16(b_3 + 1) $ and arbitrary $\mathbf{M} \in \mathbb{C}^{d_1 \times d_2}$, we have
\ea{
& \big\| \big(\mathcal{P}_T{\mathcal{Q}_{\Omega} -  \mathcal{P}_{T}}  \big) (\mathbf{M}) \big\|_{\mathcal{A},2}   
\leq 2\big( \sqrt{\tfrac{2(b_3+1)}{c_0  \tilde{K} \mathsf{R}_{\mathscr{L}}}}  \| \mathbf{M} \|_{\rm \mathcal{A},2}  + \tfrac{2(b_3+1)}{3c_0\tilde{K} \mathsf{R}_{\mathscr{L}}} \| \mathbf{M} \|_{\rm \mathcal{A},\infty}\big), 
\nonumber
}

with a probability no less than $1-N^{-b_3}$, given that $ p_{n} \geq c_0 \frac{\tilde{\mu}_{n} \mathsf{R}_{\mathscr{L}}\tilde{K}^2 }{N} {\log(N)} $ for $n \in [N]$.
\end{lemma}
\begin{proof}
See Supplementary Section VI. 
\end{proof}
\begin{lemma}
\label{lemma_6}
Suppose we have that 
\begin{align} \label{eq:Condlem6}
\frac{1}{8\sqrt{\log(N)}} \leq \min_{i\in [N]}\Big\{ \|\mathbf{A}_{i}\|_{0}\min\{\|\mathcal{P}_{U}(\mathbf{A}_{i})\|_{\rm F}^2, \|\mathcal{P}_{V}(\mathbf{A}_{i})\|_{\rm F}^2\} \Big\}.
\end{align}
Then, for $ c_0 \geq 144(b_4 + 1) $ and arbitrary $ \mathbf{M} \in T $, we have 
\ea{\label{eq:lemma6}
& \Big\| \Big(  {\mathcal{P}_T \mathcal{Q}_{\Omega}} - \mathcal{P}_T  \Big) (\mathbf{M}) \Big\|_{\rm  \mathcal{A},\infty  } 
\leq \sqrt{72} \Big( \sqrt{\tfrac{2(b_4+1)}{c_0}} \left\| \mathbf{M} \right\|_{\rm \mathcal{A}, 2 }
 + \tfrac{2(b_4+1)}{3c_0} \left\| \mathbf{M} \right\|_{\rm \mathcal{A}, \infty } \Big) \nonumber,
}
with probability at least $1-N^{-b_4+1}$, given that $ p_{n} \geq c_0 \frac{\tilde{\mu}_{n}  \tilde{K}^2}{N} {\log(N)}$ for~$n\in [N]$.
\end{lemma}
\begin{proof}
The proof is provided in Supplementary Sec. 7 of this document.
\end{proof}

In the next subsection we find the upper bound on $ \| \mathcal{P}_{T^{\perp}} (\mathbf{G}) \| $ by using the stated lemmas and defined norms in \eqref{eq:norms}.

\subsection{Upper bound derivation}
Having introduced the norms in \eqref{eq:norms} and the lemmas in Sec. \ref{sec:Some relevant lemmas}, we can now return to our main objective: upper bounding $ \| \mathcal{P}_{T^{\perp}} (\mathbf{G}) \| $.
Note that all lemmas in Sec. \ref{sec:Some relevant lemmas} hold for some universal constant $c_0$.  
For convenience, in the remainder of the proof we shall take the value of $c_0$ for which all bounds in these lemmas hold, that is, we take 
\begin{align}
c_0\geq \max \big\{ 2(b_2+1) \,,\, 16(b_3+1) \,,\, 144(b_4+1) \,,\, \tfrac{53}{3} (b_1+1) \big\}.
\end{align}
Recalling $(60)$ (in the main manuscript), we have
\begin{align}
\label{up_bound_1}
\left\| \mathcal{P}_{T^{\perp}} (\mathbf{G})  \right\| \leq \sum_{\ell\in [L]}{\left\| \mathcal{P}_{T^{\perp}}   \mathcal{Q}_{\Omega_{\ell}}  \mathcal{P}_T (\mathbf{F}_{\ell-1})  \right\|}.
\end{align}
Next, we bound each term in the right hand summation of \eqref{up_bound_1} as
\begin{align}\label{eq:final_bound_1_1}
\nonumber
\| & \mathcal{P}_{T^{\perp}} \mathcal{Q}_{\Omega_{\ell}} \mathcal{P}_T (\mathbf{F}_{\ell-1})  \|  
=\big\|  \big(\mathcal{P}_{T^{\perp}} (\mathcal{Q}_{\Omega_{\ell}} - \mathcal{I}) \mathcal{P}_T\big) (\mathbf{F}_{\ell-1})  \| \\
& \leq \big\|  \big( (\mathcal{Q}_{\Omega_{\ell}} - \mathcal{I}) \mathcal{P}_T\big) (\mathbf{F}_{\ell-1})  \|
= \left\| \left( \mathcal{Q}_{\Omega_{\ell}} - \mathcal{I} \right)  (\mathbf{F}_{\ell-1}) \right\| \nonumber\\  
& \hspace{-0.8em}\stackrel{{(a)}}{\leq} \sqrt{\tfrac{2(b_2+1)}{c_0 \tilde{K} \mathsf{R}_{\mathscr{L}}}} \| \mathbf{F}_{\ell-1} \|_{\rm \mathcal{A},2} + \tfrac{2(b_2+1)}{3c_0 \tilde{K} \mathsf{R}_{\mathscr{L}}}\| \mathbf{F}_{\ell-1} \|_{\rm \mathcal{A},\infty} \nonumber\\ 
&  \leq \frac{ \| \mathbf{F}_{\ell-1} \|_{\rm \mathcal{A},2} + \| \mathbf{F}_{\ell-1} \|_{\rm \mathcal{A},\infty}}{c_1 \sqrt{\tilde{K} \mathsf{R}_{\mathscr{L}}}}  ,
\end{align}
with probability $1-N^{-b_2}$ 
where 
in (a)  we use Lemma \ref{lemma4} and for
$$ c_1 = \min \left\{ \tfrac{3c_0 \sqrt{\tilde{K} \mathsf{R}_{\mathscr{L}}}}{2(b_2+1)}, \sqrt{\tfrac{c_0}{2(b_2+1)}} \right\}. $$ \vspace{-5pt}
Thus,
\begin{align}\label{eq:f}
\left\| \mathcal{P}_{T^{\perp}} (\mathbf{G})  \right\| \leq
\tfrac{1}{c_1\sqrt{\tilde{K} \mathsf{R}_{\mathscr{L}}}} \sum_{\ell\in [L]} \big( \| \mathbf{F}_{\ell-1} \|_{\rm \mathcal{A},2} + \| \mathbf{F}_{\ell-1} \|_{\rm \mathcal{A},\infty} \big)
\end{align}
holds with probability no less than $1-L N^{-b_2}$.
Since $\mathbf{F}_{\ell} = \Big( \mathcal{P}_{T} - \mathcal{P}_{T}\mathcal{Q}_{\Omega_{\ell}} \Big)(\mathbf{F}_{\ell-1})$, we use Lemmas \ref{lemma5} and \ref{lemma_6} to recursively bound $\|  \mathcal{P}_{T^{\perp}} \mathcal{Q}_{\Omega_{\ell}} \mathcal{P}_T (\mathbf{F}_{\ell-1})  \|$ as
\eas{
\label{mid_bound}
& \| \mathbf{F}_{\ell} \|_{\rm \mathcal{A},2} + \| \mathbf{F}_{\ell} \|_{\rm \mathcal{A},\infty} 
 \leq  \big(\sqrt{\tfrac{8(b_3+1)}{c_0}} + \sqrt{\tfrac{144(b_4+1)}{c_0}} \big) \left\| \mathbf{F}_{\ell-1} \right\|_{\rm \mathcal{A}, 2 } \\
 & + 
 \big(\tfrac{4(b_3+1)}{3c_0} + \tfrac{2\sqrt{72}(b_4+1)}{3c_0} \big) \left\| \mathbf{F}_{\ell-1}\right\|_{\rm \mathcal{A}, \infty } 
 \leq   \frac{\left\| \mathbf{F}_{\ell-1} \right\|_{\rm \mathcal{A}, 2 } +  \left\| \mathbf{F}_{\ell-1} \right\|_{\rm \mathcal{A}, \infty }}{c_2},
}
with probability no less than $1-N^{-b_3}-N^{-b_4+1}$, where 
\begin{align*}
c_2 = \min \left\{ \tfrac{1}{\sqrt{\frac{8(b_3+1)}{c_0}} + \sqrt{\frac{144(b_4+1)}{c_0}}} , \tfrac{1}{\frac{4(b_3+1)}{3c_0} + \frac{2\sqrt{72}(b_4+1)}{3c_0} } \right\} .
\end{align*}
By applying \eqref{mid_bound} multiple times, we conclude that
\begin{align}
\label{eq:PT_G_bound}
\left\| \mathcal{P}_{T^{\perp}} (\mathbf{G})  \right\| 
\leq \frac{ \| \mathbf{F}_{0} \|_{\rm \mathcal{A},2} + \| \mathbf{F}_{0} \|_{\rm \mathcal{A},\infty} }{c_1 \sqrt{\tilde{K} \mathsf{R}_{\mathscr{L}}} } \sum_{\ell\in[L]} c_2^{1-\ell},
\end{align}
with probability no less than $1-LN \sum_{i=2}^{4}N^{-b_i}$. 
We further bound $ \| \mathbf{F}_{0} \|_{\rm \mathcal{A},\infty} $ and $ \| \mathbf{F}_{0} \|_{\rm \mathcal{A},2}$ to simplify \eqref{eq:PT_G_bound}.
To bound $ \| \mathbf{F}_{0} \|_{\rm \mathcal{A}, \infty}$, we first recall that
\begin{align}\label{eq:F0_infinity}
\| \mathbf{F}_{0} \|_{\rm \mathcal{A}, \infty} = \max_{n\in[N]} \Big| 	\frac{\langle \mathbf{A}_{n} \,,\, \overbrace{\mathbf{W}_{L}^{\mathsf{H} }\mathbf{UV}^{\mathsf{H}} \mathbf{W}_{R}}^{\mathbf{F}_{0}} \, \rangle N}{\sqrt{\omega_{n}} \tilde{\mu}_{n} \tilde{K}} \Big|.
\end{align} 
In addition, we know
\begin{align}
\nonumber
\langle \mathbf{A}_{n} \,,\, & \mathbf{W}_{L}^{\mathsf{H} }\mathbf{UV}^{\mathsf{H}} \mathbf{W}_{R} \rangle   
= \langle \mathbf{U}^{\mathsf{H}}\mathbf{W}_{L}\mathbf{A}_{n} \,,\, \mathbf{V}^{\mathsf{H}}\mathbf{W}_{R} \rangle  \\ 
& = \sqrt{\omega_{n}}   \big\langle \mathbf{U}^{\mathsf{H}}\mathbf{W}_{L}\mathbf{A}_{n} \,,\, \mathbf{V}^{\mathsf{H}} \mathbf{W}_{R} \mathbf{A}_{n}^R \big\rangle  \nonumber\\
& = \sqrt{\omega_{n}}  \big\langle \mathbf{{U}}^{\mathsf{H}} \mathbf{W}_{L} \mathcal{P}_{U}(\mathbf{A}_{n}) \,,\, (\mathcal{P}_{V}(\mathbf{A}_{n}^R) \mathbf{W}_{R}^{\mathsf{H} } \mathbf{V})^{\mathsf{H}} \big\rangle.
\label{eq:UV_inf_bound_123}
\end{align}

where $ \mathbf{A}_{n}^R$ is $ d_2 \times d_2$ right diagonal version of $ \mathbf{A}_{n} $, in which for each column its diagonal element is equal to  norm-one of that column. $\mathbf{A}_{n}^R$ comes from the fact that $\mathbf{A}_{n}$s are orthonormal basis.
\begin{align}\label{eq:PU_UB}
\big\| \mathbf{{U}}^{\mathsf{H}} \mathbf{W}_{L} \mathcal{P}_{U}(\mathbf{A}_{n}) \big\|_{\rm F} 
\leq \underbrace{\|\mathbf{{U}}^{\mathsf{H}}\|}_{\leq 1}  \underbrace{\|\mathbf{W}_{L}\|_{\rm F}}_{=1}  \underbrace{\| \mathcal{P}_{U}(\mathbf{A}_{n}) \|_{\rm F}}_{\leq \sqrt{\frac{\mu_k \tilde{K}}{N}}},
\end{align} 
and 
\begin{align}\label{eq:PV_UB}
\big\| (\mathcal{P}_{V}(\mathbf{A}_{n}^R) \mathbf{W}_{R}^{\mathsf{H} } \mathbf{V})^{\mathsf{H}} \big\|_{\rm F}
&\leq \underbrace{\|\mathbf{{V}}^{\mathsf{H}}\|}_{\leq 1}  \underbrace{\|\mathbf{W}_{R}\|_{\rm F}}_{=1}  \underbrace{\| \mathcal{P}_{V}(\mathbf{A}_{n}^R) \|_{\rm F}}_{\leq \sqrt{\frac{\mu_k \tilde{K}}{N}}}.
\end{align}  
Now, if we apply the Cauchy-Schwartz inequality in \eqref{eq:UV_inf_bound_123} by using \eqref{eq:PU_UB} and \eqref{eq:PV_UB}, we obtain
\begin{align}
\Big|\langle \mathbf{A}_{n} \,,\, & \mathbf{W}_{L}^{\mathsf{H} }\mathbf{UV}^{\mathsf{H}} \mathbf{W}_{R} \rangle   \Big|
\leq \frac{\sqrt{\omega_k} \mu_k \tilde{K}}{N}.
\label{eq:F0_bound}
\end{align}
By plugging this result into \eqref{eq:F0_infinity}, we get $\| \mathbf{F}_{0} \|_{\rm \mathcal{A}, \infty} \leq 1$. For bounding $ \| \mathbf{F}_{0} \|_{\mathcal{A}, 2}$, we use
\vspace{-5pt}
\begin{align}
\| \mathbf{F}_{0} \|_{\mathcal{A}, 2}^2 &= \sum_{n \in [N]} \tfrac{N \, | \langle \mathbf{A}_{n}, \mathbf{F}_{0} \rangle |^2  }{{\omega_{n}} \tilde{\mu}_{n} \tilde{K} }
= \sum_{n \in [N]} \tfrac{\tilde{\mu}_{n}\tilde{K}}{N}\Big(\tfrac{ N \, | \langle \mathbf{A}_{n}, \mathbf{F}_{0} \rangle | }{\sqrt{\omega_{n}} \tilde{\mu}_{n} \tilde{K} }\Big)^2 \nonumber\\
& \leq \sum_{n \in [N]} \frac{\tilde{\mu}_{n}\tilde{K}}{N} 
\leq  \sum_{n \in [N]} \| \mathcal{P}_{U}(\mathbf{A}_{n}) \|_{\rm F}^2 + \| \mathcal{P}_{V}(\mathbf{A}_{n}) \|_{\rm F}^2 \label{eq:UV_2_bound}
\end{align}
 By invoking (24), we further bound $ \sum_{n \in [N]} {\| \mathcal{P}_{U}(\mathbf{A}_{n}) \|_{\rm F}^2} $ as
\begin{align}
\nonumber
\sum_{n \in [N]} &{\| \mathcal{P}_{U}(\mathbf{A}_{n}) \|_{\rm F}^2} 
\leq {\Big\| \sum_{n \in [N]} (\mathbf{A}_{n}\odot \mathbf{A}_{n}) \Big\|_{\infty\rightarrow \infty}}  \|  \mathcal{P}_{U}(\mathbf{1}) \|_{\rm F}^2 \\ & \leq \underbrace{ \sum_{n \in [N]} \|\mathbf{A}_{n}\odot \mathbf{A}_{n})\|_{\infty\rightarrow \infty}}_{\mathsf{R}_{\mathscr{L}}} \tilde{K} \leq \mathsf{R}_{\mathscr{L}} \tilde{K}.
\end{align}
A similar approach shows that $ \sum_{n \in [N]} {\| \mathcal{P}_{V}(\mathbf{A}_{n}) \|_{\rm F}^2} \leq \tilde{K} \mathsf{R}_{\mathscr{L}} $. 
Hence, $ \sum_{n \in [N]}  {\frac{| \langle \mathbf{A}_{n}, \mathbf{F}_{0} \rangle |^2 N }{{\omega_{n}} \tilde{\mu}_{n} \tilde{K} }} \leq 2\tilde{K}\mathsf{R}_{\mathscr{L}}$, or
$ 
\| \mathbf{F}_{0} \|_{\rm \mathcal{A},2}^2 \leq 2\tilde{K}\mathsf{R}_{\mathscr{L}}
$.
We now get back to \eqref{eq:PT_G_bound}:
\begin{align}
\left\| \mathcal{P}_{T^{\perp}} (\mathbf{G})  \right\| 
\leq \frac{ \sqrt{2\tilde{K}\mathsf{R}_{\mathscr{L}}} + 1 }{c_1 \sqrt{\tilde{K} \mathsf{R}_{\mathscr{L}}} } \sum_{\ell\in [L]} c_2^{1-\ell} \leq \frac{ 2\sqrt{2} }{c_1} \sum_{\ell\in [L]} c_2^{1-\ell}
\end{align}
for $ q_{n} \geq c_0 \mathsf{R}_{\mathscr{L}}^2 \frac{\tilde{\mu}_{n}}{N}\tilde{K}^2  $, or equivalently $ p_{n} \geq {c}_0 \mathsf{R}_{\mathscr{L}}^2 \frac{\tilde{\mu}_{n}}{N}\tilde{K}^2 \log(N)  $. For $ c_2 \geq 2 $ and $ c_1 \geq 12 $,  we can conclude that
\begin{align}
\label{eq:sumcl}
\left\| \mathcal{P}_{T^{\perp}} (\mathbf{G})  \right\| \leq \tfrac{ 2\sqrt{2} }{c_1} \Big(1 + \sum_{\ell=1}^{\infty} (\tfrac{1}{2})^{\ell} \Big) \leq \tfrac{ 4\sqrt{2} }{c_1} \leq \frac{1}{2},
\end{align}
with probability at least $1-LN \sum_{i=2}^{4}N^{-b_i}$. 
Therefore, if $ p_{n} \geq c_0 \mathsf{R}_{\mathscr{L}}^2 \frac{\tilde{\mu}_{n}}{N}  \tilde{K}^2  \log{(N)}$ for $n\in [N]$, with  probability no less than
$1-LN \sum_{i=1}^{4}N^{-b_i}$, matrix $ \mathbf{G} $  is a valid dual certificate. As a result, from Lemma 1, the solution of weighted lifted-structured low-rank matrix recovery problem is exact and unique, with high probability.

\bibliographystyle{elsarticle-num}
\bibliography{IEEEabrv,references}
\end{document}


\title{Supplementary material for: \\ 
	``Harmonic Retrieval Using Weighted Lifted-Structure Low-Rank Matrix Completion''
	}
	\author{Mohammad~Bokaei, Saeed~Razavikia, Arash~Amini, and Stefano Rini}
	\maketitle

\addtocounter{equation}{25}

\addtocounter{lemma}{2}

\section{An Upper Bound on $ \| \mathcal{P}_{T^{\perp}} (\mathbf{G}) \| $}\label{sec:up_bound}

For simplicity in referencing, it should be noted that the number of equations here starts after the last equation number of the original paper, i.e., (25).

\subsection{Some relevant lemmas}
\label{sec:Some relevant lemmas}
We begin by defining the following two norms for the matrix $ \mathbf{M} \in \mathbb{C}^{d_1 \times d_2}$ for a given set of lifting basis   $\{\mathbf{A}_{n}\}_{n \in [N]}$: 
\eas{
\| \mathbf{M} \| _{\mathcal{A},\infty} 
& := \max_{n\in [N]} {\bigg|  \frac{{N} \langle\mathbf{A}_{n},\mathbf{M} \rangle }{\tilde{K} \tilde{\mu}_{n} \sqrt{\omega_{n}}  } \bigg|}  
\label{def_norm_max}, \\
 \| \mathbf{M} \| _{\mathcal{A},2} 
& :=  \sqrt{ \sum_{n\in [N]} {\frac{| {N} \langle \mathbf{A}_{n}, \mathbf{M} \rangle |^2  }{\tilde{K} \tilde{\mu}_{n} \omega_{n} }} },
\label{def_norm_2} 
}{\label{eq:norms}}
where we have defined $\omega_n:=\|\mathbf{A}_n\|_0$.
We now state three inequalities regarding the  norms in \eqref{eq:norms}: Lemma  \ref{lemma4}, Lemma \ref{lemma5}, and Lemma \ref{lemma_6}.
Generally, these proofs rely on  matrix concentration inequalities in \cite[Appendix~A]{chen2014robust}.

\begin{lemma}
\label{lemma4}
Suppose $ \mathbf{M}$ is a complex-valued ${d_1 \times d_2}$ matrix. If  $ p_{n} \geq c_0 \frac{\tilde{\mu}_{n} \tilde{K}^2 \log(N)}{N} $ for all $n \in [N]$, then
\ea{
& \big\| \big({\mathcal{Q}_{\Omega} -  \mathcal{I} }  \big) \mathbf{M} \big\|  \\
& \quad \quad \quad \leq  \sqrt{\tfrac{2(b_2+1)}{c_0  \tilde{K} \mathsf{R}_{\mathscr{L}}}}  \| \mathbf{M} \|_{\rm \mathcal{A},2}  + \tfrac{2(b_2+1)}{3c_0\tilde{K} \mathsf{R}_{\mathscr{L}}} \| \mathbf{M} \|_{\rm \mathcal{A},\infty}, 
\nonumber
}
holds with a probability at least $1-N^{-b_2}$, where $ c_0 \geq 2(b_2 + 1)$ and ${\mathsf{R}}_{\mathscr{L}} = \sum_{n \in [N]}  \big\|\mathbf{A}_{n} \odot
    \mathbf{A}_{n}\big\|_{\infty \rightarrow \infty}$.
\end{lemma}
\begin{proof}
The proof is provided in Section \ref{lem4_proof}. 
\end{proof}
We further control $ \|.\|_{\mathcal{A}, 2} $ and $ \|.\|_{\mathcal{A}, \infty} $ norms of $ \big( {\mathcal{P}_T \mathcal{Q}_{\Omega}} - \mathcal{P}_T  \big) $ in the next two lemmas.

\begin{lemma}
\label{lemma5}
For $ c_0 \geq 16(b_3 + 1) $ and arbitrary $\mathbf{M} \in \mathbb{C}^{d_1 \times d_2}$, we have
\ea{
& \big\| \big(\mathcal{P}_T{\mathcal{Q}_{\Omega} -  \mathcal{P}_{T}}  \big) (\mathbf{M}) \big\|_{\mathcal{A},2}   \label{eq:lemma5} \\
& \quad \quad \quad \leq 2\big( \sqrt{\tfrac{2(b_3+1)}{c_0  \tilde{K} \mathsf{R}_{\mathscr{L}}}}  \| \mathbf{M} \|_{\rm \mathcal{A},2}  + \tfrac{2(b_3+1)}{3c_0\tilde{K} \mathsf{R}_{\mathscr{L}}} \| \mathbf{M} \|_{\rm \mathcal{A},\infty}\big), 
\nonumber
}
with a probability no less than $1-N^{-b_3}$, given that $ p_{n} \geq c_0 \frac{\tilde{\mu}_{n} \mathsf{R}_{\mathscr{L}}\tilde{K}^2 }{N} {\log(N)} $ for $n \in [N]$.
\end{lemma}
\begin{proof}
For the proof see Section \ref{lem2_proof}. 
\end{proof}
\begin{lemma}
\label{lemma_6}
Suppose we have that { \change
\begin{align}
\label{eq:Condlem6}
\frac{1}{8\sqrt{\log(N)}}\leq\min_{i\in [N]}\Big\{ \|\mathbf{A}_{i}\|_{0}\min\{\|\mathcal{P}_{U}(\mathbf{A}_{i})\|_{\rm F}^2, \|\mathcal{P}_{V}(\mathbf{A}_{i})\|_{\rm F}^2\} \Big\}.
\end{align}}
Then, for $ c_0 \geq 144(b_4 + 1) $ and arbitrary $ \mathbf{M} \in T $, we have 
\ea{\label{eq:lemma6}
& \Big\| \Big(  {\mathcal{P}_T \mathcal{Q}_{\Omega}} - \mathcal{P}_T  \Big) (\mathbf{M}) \Big\|_{\rm  \mathcal{A},\infty  }  \\
& \quad \quad \quad \leq \sqrt{72} \Big( \sqrt{\tfrac{2(b_4+1)}{c_0}} \left\| \mathbf{M} \right\|_{\rm \mathcal{A}, 2 }
 + \tfrac{2(b_4+1)}{3c_0} \left\| \mathbf{M} \right\|_{\rm \mathcal{A}, \infty } \Big) \nonumber,
}
with probability at least $1-N^{-b_4+1}$, given that $ p_{n} \geq c_0 \frac{\tilde{\mu}_{n}  \tilde{K}^2}{N} {\log(N)}$ for $n\in [N] $.
\end{lemma}
\begin{proof}
The proof is provided in Sec. \ref{lem6_proof}
of this document.
\end{proof}

In the following subsection, we find the upper bound on $ \| \mathcal{P}_{T^{\perp}} (\mathbf{G}) \| $ by using the stated lemmas and defined norms in \eqref{eq:norms}.

\subsection{Upper bound derivation}
Having introduced the norms in \eqref{eq:norms} and the lemmas in Sec. \ref{sec:Some relevant lemmas}, we can now return to our main objective: upper bounding $ \| \mathcal{P}_{T^{\perp}} (\mathbf{G}) \| $.
Note that all lemmas in Sec. \ref{sec:Some relevant lemmas} hold for some universal constant $c_0$. For convenience, in the remainder of the proof, we shall take the value of $c_0$ for which all bounds in these lemmas hold, that is, we take 
\begin{align}
c_0\geq \max \big\{ 2(b_2+1) \,,\, 16(b_3+1) \,,\, 144(b_4+1) \,,\, \tfrac{53}{3} (b_1+1) \big\}.
\end{align}
Recalling $(A.15)$ (in the main manuscript), we have
\begin{align}
\label{up_bound_1}
\left\| \mathcal{P}_{T^{\perp}} (\mathbf{G})  \right\| \leq \sum_{\ell\in [L]}{\left\| \mathcal{P}_{T^{\perp}}   \mathcal{Q}_{\Omega_{\ell}}  \mathcal{P}_T (\mathbf{F}_{\ell-1})  \right\|}.
\end{align}
Next, we bound each term in the right-hand summation of \eqref{up_bound_1} as
\begin{align}\label{eq:final_bound_1_1}
\nonumber
\| & \mathcal{P}_{T^{\perp}} \mathcal{Q}_{\Omega_{\ell}} \mathcal{P}_T (\mathbf{F}_{\ell-1})  \|  
=\big\|  \big(\mathcal{P}_{T^{\perp}} (\mathcal{Q}_{\Omega_{\ell}} - \mathcal{I}) \mathcal{P}_T\big) (\mathbf{F}_{\ell-1})  \| \\
& \leq \big\|  \big( (\mathcal{Q}_{\Omega_{\ell}} - \mathcal{I}) \mathcal{P}_T\big) (\mathbf{F}_{\ell-1})  \|
= \left\| \left( \mathcal{Q}_{\Omega_{\ell}} - \mathcal{I} \right)  (\mathbf{F}_{\ell-1}) \right\| \nonumber\\  
& \hspace{-0.8em}\stackrel{{(a)}}{\leq} \sqrt{\tfrac{2(b_2+1)}{c_0 \tilde{K} \mathsf{R}_{\mathscr{L}}}} \| \mathbf{F}_{\ell-1} \|_{\rm \mathcal{A},2} + \tfrac{2(b_2+1)}{3c_0 \tilde{K} \mathsf{R}_{\mathscr{L}}}\| \mathbf{F}_{\ell-1} \|_{\rm \mathcal{A},\infty} \nonumber\\ 
&  \leq \frac{ \| \mathbf{F}_{\ell-1} \|_{\rm \mathcal{A},2} + \| \mathbf{F}_{\ell-1} \|_{\rm \mathcal{A},\infty}}{c_1 \sqrt{\tilde{K} \mathsf{R}_{\mathscr{L}}}}  ,
\end{align}
with probability $1-N^{-b_2}$ 
where 
in (a)  we use Lemma \ref{lemma4} and for
$$ c_1 = \min \left\{ \tfrac{3c_0 \sqrt{\tilde{K} \mathsf{R}_{\mathscr{L}}}}{2(b_2+1)}, \sqrt{\tfrac{c_0}{2(b_2+1)}} \right\}. $$ 
Thus,
\begin{align}\label{eq:f}
\left\| \mathcal{P}_{T^{\perp}} (\mathbf{G})  \right\| \leq
\tfrac{1}{c_1\sqrt{\tilde{K} \mathsf{R}_{\mathscr{L}}}} \sum_{\ell\in [L]} \big( \| \mathbf{F}_{\ell-1} \|_{\rm \mathcal{A},2} + \| \mathbf{F}_{\ell-1} \|_{\rm \mathcal{A},\infty} \big)
\end{align}
holds with probability no less than $1-L N^{-b_2}$.

Since $\mathbf{F}_{\ell} = \Big( \mathcal{P}_{T} - \mathcal{P}_{T}\mathcal{Q}_{\Omega_{\ell}} \Big)(\mathbf{F}_{\ell-1})$, we use Lemmas \ref{lemma5} and \ref{lemma_6} to recursively bound $\|  \mathcal{P}_{T^{\perp}} \mathcal{Q}_{\Omega_{\ell}} \mathcal{P}_T (\mathbf{F}_{\ell-1})  \|$ as
\eas{
\label{mid_bound}
& \| \mathbf{F}_{\ell} \|_{\rm \mathcal{A},2} + \| \mathbf{F}_{\ell} \|_{\rm \mathcal{A},\infty} \\
& \leq  \big(\sqrt{\tfrac{8(b_3+1)}{c_0}} + \sqrt{\tfrac{144(b_4+1)}{c_0}} \big) \left\| \mathbf{F}_{\ell-1} \right\|_{\rm \mathcal{A}, 2 } \\
 & \quad \quad \quad + \big(\tfrac{4(b_3+1)}{3c_0} + \tfrac{2\sqrt{72}(b_4+1)}{3c_0} \big) \left\| \mathbf{F}_{\ell-1}\right\|_{\rm \mathcal{A}, \infty } \nonumber \\
& \leq   \frac{\left\| \mathbf{F}_{\ell-1} \right\|_{\rm \mathcal{A}, 2 } +  \left\| \mathbf{F}_{\ell-1} \right\|_{\rm \mathcal{A}, \infty }}{c_2},
}
with probability no less than $1-N^{-b_3}-N^{-b_4+1}$, where 
\begin{align*}
c_2 = \min \left\{ \tfrac{1}{\sqrt{\frac{8(b_3+1)}{c_0}} + \sqrt{\frac{144(b_4+1)}{c_0}}} , \tfrac{1}{\frac{4(b_3+1)}{3c_0} + \frac{2\sqrt{72}(b_4+1)}{3c_0} } \right\} .
\end{align*}
By applying \eqref{mid_bound} multiple times, we conclude that
\begin{align}
\label{eq:PT_G_bound}
\left\| \mathcal{P}_{T^{\perp}} (\mathbf{G})  \right\| 
\leq \frac{ \| \mathbf{F}_{0} \|_{\rm \mathcal{A},2} + \| \mathbf{F}_{0} \|_{\rm \mathcal{A},\infty} }{c_1 \sqrt{\tilde{K} \mathsf{R}_{\mathscr{L}}} } \sum_{\ell\in[L]} c_2^{1-\ell}
\end{align}
with probability no less than $1-LN \sum_{i=2}^{4}N^{-b_i}$. 
We further bound $ \| \mathbf{F}_{0} \|_{\rm \mathcal{A},\infty} $ and $ \| \mathbf{F}_{0} \|_{\rm \mathcal{A},2}$ to simplify \eqref{eq:PT_G_bound}.

To bound $ \| \mathbf{F}_{0} \|_{\rm \mathcal{A}, \infty}$, we first recall that
\begin{align}\label{eq:F0_infinity}
\| \mathbf{F}_{0} \|_{\rm \mathcal{A}, \infty} = \max_{n\in[N]} \Big| 	\frac{\langle \mathbf{A}_{n} \,,\, \overbrace{\mathbf{W}_{L}^{\mathsf{H} }\mathbf{UV}^{\mathsf{H}} \mathbf{W}_{R}}^{\mathbf{F}_{0}} \, \rangle N}{\sqrt{\omega_{n}} \tilde{\mu}_{n} \tilde{K}} \Big|.
\end{align} 
In addition, we know
\begin{align}
\nonumber
\langle \mathbf{A}_{n} \,,\, & \mathbf{W}_{L}^{\mathsf{H} }\mathbf{UV}^{\mathsf{H}} \mathbf{W}_{R} \rangle   
= \langle \mathbf{U}^{\mathsf{H}}\mathbf{W}_{L}\mathbf{A}_{n} \,,\, \mathbf{V}^{\mathsf{H}}\mathbf{W}_{R} \rangle  \\ 
& = \sqrt{\omega_{n}}   \big\langle \mathbf{U}^{\mathsf{H}}\mathbf{W}_{L}\mathbf{A}_{n} \,,\, \mathbf{V}^{\mathsf{H}} \mathbf{W}_{R} \mathbf{A}_{n}^R \big\rangle  \nonumber\\
& = \sqrt{\omega_{n}}  \big\langle \mathbf{{U}}^{\mathsf{H}} \mathbf{W}_{L} \mathcal{P}_{U}(\mathbf{A}_{n}) \,,\, (\mathcal{P}_{V}(\mathbf{A}_{n}^R) \mathbf{W}_{R}^{\mathsf{H} } \mathbf{V})^{\mathsf{H}} \big\rangle.
\label{eq:UV_inf_bound_123}
\end{align}
{\change
where $ \mathbf{A}_{n}^R$ is $ d_2 \times d_2$ right diagonal version of $ \mathbf{A}_{n} $, in which for each column its diagonal element is equal to norm-one of that column. $\mathbf{A}_{n}^R$ comes from the fact that $\mathbf{A}_{n}$s are orthonormal basis.}
\begin{align}\label{eq:PU_UB}
\big\| \mathbf{{U}}^{\mathsf{H}} \mathbf{W}_{L} \mathcal{P}_{U}(\mathbf{A}_{n}) \big\|_{\rm F} 
\leq \underbrace{\|\mathbf{{U}}^{\mathsf{H}}\|}_{\leq 1}  \underbrace{\|\mathbf{W}_{L}\|_{\rm F}}_{=1}  \underbrace{\| \mathcal{P}_{U}(\mathbf{A}_{n}) \|_{\rm F}}_{\leq \sqrt{\frac{\mu_k \tilde{K}}{N}}},
\end{align} 
and 
\begin{align}\label{eq:PV_UB}
\big\| (\mathcal{P}_{V}(\mathbf{A}_{n}^R) \mathbf{W}_{R}^{\mathsf{H} } \mathbf{V})^{\mathsf{H}} \big\|_{\rm F}
&\leq \underbrace{\|\mathbf{{V}}^{\mathsf{H}}\|}_{\leq 1}  \underbrace{\|\mathbf{W}_{R}\|_{\rm F}}_{=1}  \underbrace{\| \mathcal{P}_{V}(\mathbf{A}_{n}^R) \|_{\rm F}}_{\leq \sqrt{\frac{\mu_k \tilde{K}}{N}}}.
\end{align}  
Now, if we apply the Cauchy-Schwartz inequality in \eqref{eq:UV_inf_bound_123} by using \eqref{eq:PU_UB} and \eqref{eq:PV_UB}, we obtain
\begin{align}
\Big|\langle \mathbf{A}_{n} \,,\, & \mathbf{W}_{L}^{\mathsf{H} }\mathbf{UV}^{\mathsf{H}} \mathbf{W}_{R} \rangle   \Big|
\leq \frac{\sqrt{\omega_k} \mu_k \tilde{K}}{N}.
\label{eq:F0_bound}
\end{align}
By plugging this result into \eqref{eq:F0_infinity}, we get $\| \mathbf{F}_{0} \|_{\rm \mathcal{A}, \infty} \leq 1$. For bounding $ \| \mathbf{F}_{0} \|_{\mathcal{A}, 2}$, we use
\begin{align}
\| \mathbf{F}_{0} \|_{\mathcal{A}, 2}^2 &= \sum_{n \in [N]} \tfrac{N \, | \langle \mathbf{A}_{n}, \mathbf{F}_{0} \rangle |^2  }{{\omega_{n}} \tilde{\mu}_{n} \tilde{K} }
= \sum_{n \in [N]} \tfrac{\tilde{\mu}_{n}\tilde{K}}{N}\Big(\tfrac{ N \, | \langle \mathbf{A}_{n}, \mathbf{F}_{0} \rangle | }{\sqrt{\omega_{n}} \tilde{\mu}_{n} \tilde{K} }\Big)^2 \nonumber\\
& \leq \sum_{n \in [N]} \frac{\tilde{\mu}_{n}\tilde{K}}{N} 
\leq  \sum_{n \in [N]} \| \mathcal{P}_{U}(\mathbf{A}_{n}) \|_{\rm F}^2 + \| \mathcal{P}_{V}(\mathbf{A}_{n}) \|_{\rm F}^2 \label{eq:UV_2_bound}
\end{align}
{ \change By invoking (21), we further bound $ \sum_{n \in [N]} {\| \mathcal{P}_{U}(\mathbf{A}_{n}) \|_{\rm F}^2} $ as
\begin{align}
\nonumber
\sum_{n \in [N]} &{\| \mathcal{P}_{U}(\mathbf{A}_{n}) \|_{\rm F}^2} 
\leq {\Big\| \sum_{n \in [N]} (\mathbf{A}_{n}\odot \mathbf{A}_{n}) \Big\|_{\infty\rightarrow \infty}}  \|  \mathcal{P}_{U}(\mathbf{1}) \|_{\rm F}^2 \\ & \leq \underbrace{ \sum_{n \in [N]} \|\mathbf{A}_{n}\odot \mathbf{A}_{n})\|_{\infty\rightarrow \infty}}_{\mathsf{R}_{\mathscr{L}}} \tilde{K} \leq \mathsf{R}_{\mathscr{L}} \tilde{K}.
\end{align}}
A similar approach shows that $ \sum_{n \in [N]} {\| \mathcal{P}_{V}(\mathbf{A}_{n}) \|_{\rm F}^2} \leq \tilde{K} \mathsf{R}_{\mathscr{L}} $. 
Hence, $ \sum_{n \in [N]}  {\frac{| \langle \mathbf{A}_{n}, \mathbf{F}_{0} \rangle |^2 N }{{\omega_{n}} \tilde{\mu}_{n} \tilde{K} }} \leq 2\tilde{K}\mathsf{R}_{\mathscr{L}}$, or
\begin{align}
\| \mathbf{F}_{0} \|_{\rm \mathcal{A},2}^2 \leq 2\tilde{K}\mathsf{R}_{\mathscr{L}}.
\end{align}
We now get back to \eqref{eq:PT_G_bound}:
\begin{align}
\left\| \mathcal{P}_{T^{\perp}} (\mathbf{G})  \right\| 
\leq \frac{ \sqrt{2\tilde{K}\mathsf{R}_{\mathscr{L}}} + 1 }{c_1 \sqrt{\tilde{K} \mathsf{R}_{\mathscr{L}}} } \sum_{\ell\in [L]} c_2^{1-\ell} \leq \frac{ 2\sqrt{2} }{c_1} \sum_{\ell\in [L]} c_2^{1-\ell}
\end{align}
for $ q_{n} \geq c_0 \mathsf{R}_{\mathscr{L}}^2 \frac{\tilde{\mu}_{n}}{N}\tilde{K}^2  $, or equivalently $ p_{n} \geq {c}_0 \mathsf{R}_{\mathscr{L}}^2 \frac{\tilde{\mu}_{n}}{N}\tilde{K}^2 \log(N)  $. For $ c_2 \geq 2 $ and $ c_1 \geq 12 $,  we can conclude that
\begin{align}
\label{eq:sumcl}
\left\| \mathcal{P}_{T^{\perp}} (\mathbf{G})  \right\| \leq \tfrac{ 2\sqrt{2} }{c_1} \Big(1 + \sum_{\ell=1}^{\infty} (\tfrac{1}{2})^{\ell} \Big) \leq \tfrac{ 4\sqrt{2} }{c_1} \leq \frac{1}{2},
\end{align}
with probability at least $1-LN \sum_{i=2}^{4}N^{-b_i}$. Therefore, if $ p_{n} \geq c_0 \mathsf{R}_{\mathscr{L}}^2 \frac{\tilde{\mu}_{n}}{N}  \tilde{K}^2  \log{(N)}$ for $n\in [N]$, with  probability no less than
$1-LN \sum_{i=1}^{4}N^{-b_i}$, matrix $ \mathbf{G} $  is a valid dual certificate. As a result, from Lemma 1, the solution of the weighted lifted-structured low-rank matrix recovery problem is exact and unique, with high probability.

\section{Proof of Theorem 2}
	\label{noisy_recovery_proof}
	Similar to the proof of Theorem 1, we know that (19) implies that (A.9)  holds with a probability no less than $1-LN\sum_{i=1}^{4}N^{-b_i}$; besides, there exists a $\mathbf{G}$ such that (A.10-A12) hold (with the same probability).  Let $\widehat{\mathbf{y}}_{e} = \mathbf{y} + \Delta $ be the solution to (15), where $ \mathbf{y} $ is the ground-truth. 
	In the following, for $\Delta_{\rm H} = \mathscr{L}(\Delta)$ we bound 
	$ \| \Delta_{\rm H} \|_{\rm F}  $ by separating it into $\Omega$ and $\Omega^{\perp}$ components. 
	First, we bound $ \|\mathcal{A}_{\Omega} (\Delta_{\rm H})   \|_{\rm F} $ as 
	\begin{align} 
 	\nonumber
 	\|&  \mathcal{A}_{\Omega}{\big(\Delta_{\rm H} \big)}   \|_{\rm F} =  \|  \mathcal{A}_{\Omega}{\big(\mathscr{L}(\widehat{\mathbf{y}}_{e}) - \mathscr{L}(\mathbf{y})\big)}   \|_{\rm F} \\ 
	& \leq \|  \mathcal{A}_{\Omega} {\big( \mathscr{L}(\widehat{\mathbf{y}}_{e} -\mathbf{y}_{n}) \big)  } \|_{\rm F} + \| \mathcal{A}_{\Omega} { \big( \mathscr{L}(\mathbf{y} - \mathbf{y}_{n}) \big) } \|_{\rm F}.\label{eq:UBAOmega}
	\end{align}
	As each element of a vector $\mathbf{x}$ is repeated at most $\min({d}_1,{d}_2)$ times in $\mathscr{L}(\mathbf{x})$, we know that 
	\begin{align}
	\|\mathcal{A}_{\Omega}\big(\mathscr{L}(\mathbf{x})\big) \|_{\rm F} \leq \tfrac{\sqrt{\min({d}_1, {d}_2)}}{\min_n p_n}  \|\mathcal{P}_{\Omega}(\mathbf{x})\|_{\rm F}.
	\end{align}
	Thus, we can rewrite \eqref{eq:UBAOmega} as
	\begin{align}
	\|  \mathcal{A}_{\Omega}&{\big(\Delta_{\rm H} \big)}   \|_{\rm F} \leq \tfrac{\sqrt{\min({d}_1,{d}_2)}}{\min_n p_n} \Big( \underbrace{\|\mathcal{P}_{\Omega}(\widehat{\mathbf{y}}_{e} -\mathbf{y}_{n})\|_{\rm F} }_{\sqrt{M}\leq \eta} \nonumber\\ \label{eq:UAOmegaDelta}
	& + \underbrace{ \|\mathcal{P}_{\Omega}(\mathbf{y} - \mathbf{y}_{n})\|_{\rm F}}_{\leq \sqrt{M}\eta}\Big)
	\leq 2\sqrt{M}\eta \tfrac{\sqrt{\min({d}_1,\Tilde{d_2})}}{\min_n p_n}.
	\end{align}

	Next, let us define $f(\mathbf{X}) = \|\mathbf{W}_{L} \mathbf{X} \mathbf{W}_{R}^{\mathsf{H}}\|_{*}$ and $\mathbf{H} = \mathcal{A}_{\Omega^{\perp}}\big( \Delta_{\rm H} \big)$; hence, $\mathcal{A}_{\Omega}(\mathbf{H}) = \mathcal{A}^{\perp}(\mathbf{H}) = 0$. Since $ \widehat{\mathbf{y}}_{e}$ is the solution of (15), we shall have
		\begin{align}
	\nonumber
	f\big(  \mathscr{L}{(\mathbf{y})} \big)  &\geq f\big( \mathscr{L}{( \widehat{\mathbf{y}}_{e})}  \big)
	= f\big( \mathscr{L} (\mathbf{y}) + \Delta_{\rm H} \big) \nonumber\\
	&\geq f\big( \mathscr{L} (\mathbf{y}) + \mathbf{H} \big) - f\big( \mathcal{A}_{\Omega}(\Delta_{\rm H}) \big) .
	\label{eq:noisy_sub_grad}
	\end{align}
	Therefore,
	\begin{align} \label{eq:UpperfyA}
	f\big( \mathscr{L} (\mathbf{y}) + \mathbf{H} \big)
	\leq f\big(  \mathscr{L}{(\mathbf{y})} \big) + f\big( \mathcal{A}_{\Omega}(\Delta_{\rm H}) \big).
	\end{align}
	Besides, for $\mathbf{H} \neq 0$,
 	similar to \eqref{eq:longLemma1} and \eqref{eq:Nuclear_F_F} in the proof of Lemma 1,
 	we have that
	\begin{align}\label{eq:LowerfyA_1}
	f\big( \mathscr{L}(\mathbf{y}) + \mathbf{H}\big) \geq f\big( \mathscr{L}(\mathbf{y})\big) + \frac{\|\mathcal{P}_{T^{\perp}}(\mathbf{H})\|_{*}}{2} -  \frac{\|\mathcal{P}_{T}(\mathbf{H})\|_{\rm F}}{5 \|\mathcal{Q}_{\Omega}\|},
	\end{align}
	and
		\begin{align}\label{eq:LowerfyA_2}
\|\mathcal{P}_{T^{\perp}}(\mathbf{H})\|_{*} \geq \|\mathcal{P}_{T^{\perp}}(\mathbf{H})\|_{\rm F} \geq \frac{\|\mathcal{P}_{T}(\mathbf{H})\|_{\rm F}}{2 \|\mathcal{Q}_{\Omega}\|}
	\end{align}
Therefore, we can conclude that
	\begin{align}\label{eq:LowerfyA}
	f\big( \mathscr{L}(\mathbf{y}) + \mathbf{H}\big) \geq f\big( \mathscr{L}(\mathbf{y})\big) +\frac{1}{10}\|\mathcal{P}_{T^{\perp}} (\mathbf{H})\|_{\rm F}.
	\end{align}
Now, by combining \eqref{eq:UpperfyA} and \eqref{eq:LowerfyA}, we conclude that $ \|\mathcal{P}_{T^{\perp}} (\mathbf{H})\|_{\rm F} \leq 10 f\big( \mathcal{A}_{\Omega}(\Delta_{\rm H}) \big)$, and 	
\begin{align}
\label{eq:Pt(H)}
\|\mathcal{P}_{T} (\mathbf{H})\|_{\rm F}
\leq 
2\|\mathcal{Q}_{\Omega}\| \|\mathcal{P}_{T^{\perp}} (\mathbf{H})\|_{\rm F}
\leq 
20 \|\mathcal{Q}_{\Omega}\| f\big( \mathcal{A}_{\Omega}(\Delta_{\rm H}) \big).
\end{align}
Hence,
\begin{align}
\label{eq:H_bound}
\| \mathbf{H} \|_{\rm F} &= \| \mathcal{P}_{T}(\mathbf{H}) + \mathcal{P}_{T^{\perp}} (\mathbf{H}) \|_{\rm F} 
\nonumber\\
&\leq 10\big(1+2\|\mathcal{Q}_{\Omega}\| \big) f\big( \mathcal{A}_{\Omega}(\Delta_{\rm H}) \big).
\end{align}
	We further bound $f\big( \mathcal{A}_{\Omega}(\Delta_{\rm H}) \big)$ as
\begin{align}
f\big( & \mathcal{A}_{\Omega}(\Delta_{\rm H}) \big) = \| \mathbf{W}_L  \mathcal{A}_{\Omega}(\Delta_{\rm H}) \mathbf{W}_{R}^{\mathsf{H}} \|_{*} \nonumber\\
& \leq \Big(\underbrace{{\rm rank}\big( \mathbf{W}_L  \mathcal{A}_{\Omega}(\Delta_{\rm H}) \mathbf{W}_{R}^{\mathsf{H}} \big) }_{\leq \min({d}_1,{d}_2)} \Big)^{\frac{1}{2}} \, 
\| \mathbf{W}_L  \mathcal{A}_{\Omega}(\Delta_{\rm H}) \mathbf{W}_{R}^{\mathsf{H}} \|_{\rm F}\nonumber\\
& \leq \sqrt{\min({d}_1,{d}_2)}  \underbrace{\| \mathbf{W}_L \|_{\rm F} }_{=1}  \| \mathcal{A}_{\Omega}(\Delta_{\rm H}) \|_{\rm F}     \underbrace{\| \mathbf{W}_{R}^{\mathsf{H}} \|_{\rm F}}_{=1}\nonumber\\
&\leq 2\sqrt{M}\eta \frac{\min({d}_1,{d}_2)}{\min_{n} p_n}.
\end{align}	
This shows that
\begin{align}\label{eq:AomegaPerp_final}
\| \mathcal{A}_{\Omega^{\perp}}(\Delta_{\rm H}) \|_{\rm F} &= \| \mathbf{H} \|_{\rm F} \nonumber\\
&\leq 20\sqrt{M}\eta \big(1+2\|\mathcal{Q}_{\Omega}\| \big) \tfrac{\min({d}_1,{d}_2)}{\min_{n} p_n}.
\end{align}

Finally, we combine \eqref{eq:UAOmegaDelta} and \eqref{eq:AomegaPerp_final} to obtain
 \begin{align}\label{case_1_bound}
	\nonumber
	\| \Delta_{\rm H} \|_{\rm F} & \leq \| \mathcal{A}_{\Omega}( \Delta_{\rm H} ) \|_{\rm F} + \| \mathcal{A}_{\Omega^{\perp}}( \Delta_{\rm H} ) \|_{\rm F}  \\ \nonumber
	&\leq  \sqrt{M}\eta \big(22+40\|\mathcal{Q}_{\Omega}\| \big) \tfrac{\min({d}_1,{d}_2)}{\min_{n} p_n} \\ 
	& \leq 102\sqrt{M}\eta \tfrac{\min({d}_1,{d}_2)}{\min_{n} p_n^2},
	\end{align}
	where the \eqref{case_1_bound} follows from $\|\mathcal{Q}_{\Omega} \| \leq \|\mathcal{A}_{\Omega}\| + \| \mathcal{A}^{\perp} \| \leq \tfrac{1}{\min_n p_n} + 1$.

\section{ Prof of Lemma 1}
Let $\widehat{\mathbf{M}}\neq \mathscr{L}(\mathbf{y}) = \mathbf{Y}$ be a feasible solution to (A.8) according to $\mathcal{Q}_{\Omega}(\widehat{\mathbf{M}}) = \mathcal{Q}_{\Omega}(\mathbf{Y} ).$ This implies that 
$\mathbf{H} = \widehat{\mathbf{M}} - \mathbf{Y}$,
 satisfies $\mathcal{A}_{\Omega}(\mathbf{H})=\mathcal{A}^{\perp}(\mathbf{H}) =0$.
Let us represent the singular value decomposition (SVD) of $\mathcal{P}_{T^{\perp}}{(\mathbf{H})}$ by $\mathbf{U}_{\mathbf{H}} \mathbf{\Sigma}_{\mathbf{H}} \mathbf{V}_\mathbf{H}^{\mathsf{H}}$. Further, we define $\mathbf{B}_0 = \mathcal{P}_{T^{\perp}}(\mathbf{U}_\mathbf{H} \mathbf{V}_\mathbf{H}^{\mathsf{H}})$. We can now write that
\begin{align}
\left\langle  \mathbf{B}_0 \,,\, \mathbf{H} \right\rangle
&= \big\langle  \mathbf{U}_{\mathbf{H}} \mathbf{V}_{\mathbf{H}}^{\mathsf{H}} \,,\, \underbrace{\mathcal{P}_{T^{\perp}}{(\mathbf{H})}}_{\mathbf{U}_{\mathbf{H}} \mathbf{\Sigma}_{\mathbf{H}} \mathbf{V}_\mathbf{H}^{\mathsf{H}}} \big\rangle 
= {\rm tr}(\mathbf{\Sigma}_{\mathbf{H}}) = \| \mathcal{P}_{T^{\perp}}{(\mathbf{H})} \|_{*}.
\end{align}
It is also evident that $\|\mathbf{B}_0\| = \|\mathcal{P}_{T^{\perp}}(\mathbf{U}_\mathbf{H} \mathbf{V}_\mathbf{H}^{\mathsf{H}})\| \leq 1$. As $ \mathbf{B}_{0}  \in T^{\perp}$, we conclude that $ \mathbf{W}_{L}^{\mathsf{H}} \mathbf{U}\mathbf{V}^{\mathsf{H}} \mathbf{W}_{R} + \mathbf{B}_{0} $ is  in the sub-gradient of the $ f(\mathbf{Y}) = \| \mathbf{W}_{L} \mathbf{Y} \mathbf{W}_{R}^{\mathsf{H}} \|_{*} $ at $\mathbf{Y} $, where $\mathbf{U}$ and $\mathbf{V}$ are the unitary matrices in the SVD of $\mathbf{W}_{L} \mathbf{Y} \mathbf{W}_{R}^{\mathsf{H}}$. 
Using the property of the elements in the sub-gradient, we know that
\begin{align}
f(\underbrace{\mathbf{Y}+\mathbf{H}}_{\widehat{\mathbf{M}}}) \geq&  \, f(\mathbf{Y}) + \langle \mathbf{W}_{L}^{\mathsf{H}} \mathbf{U}\mathbf{V}^{\mathsf{H}} \mathbf{W}_{R} + \mathbf{B}_{0} \,,\, \mathbf{H}\rangle \nonumber\\
 =& f(\mathbf{Y}) +  \underbrace{ \langle \mathbf{G} \,,\, \mathbf{H}\rangle}_{=0} + \underbrace{\langle \mathbf{B}_{0} \,,\, \mathbf{H} \rangle}_{=\| \mathcal{P}_{T^{\perp}}{(\mathbf{H})} \|_{*}}
 \nonumber\\
&- \langle \mathbf{G} - \mathbf{W}_{L}^{\mathsf{H}} \mathbf{U}\mathbf{V}^{\mathsf{H}} \mathbf{W}_{R}  \,,\, \mathbf{H}\rangle \nonumber\\
=& f(\mathbf{Y})+ \| \mathcal{P}_{T^{\perp}}{(\mathbf{H})} \|_{*} -  \langle \mathcal{P}_{T^{\perp}}(\mathbf{G})   \,,\, \mathbf{H}\rangle \nonumber\\
& - \langle \underbrace{ \mathcal{P}_{T}(\mathbf{G}) - \mathbf{W}_{L}^{\mathsf{H}} \mathbf{U}\mathbf{V}^{\mathsf{H}} \mathbf{W}_{R} }_{= \mathcal{P}_{T}(\mathbf{G} - \mathbf{W}_{L}^{\mathsf{H}} \mathbf{U}\mathbf{V}^{\mathsf{H}} \mathbf{W}_{R})}  \,,\, \mathbf{H}\rangle \nonumber\\
\geq & f(\mathbf{Y}) + \| \mathcal{P}_{T^{\perp}}{(\mathbf{H})} \|_{*} - \underbrace{\| \mathcal{P}_{T^{\perp}}(\mathbf{G}) \|}_{\leq \frac{1}{2}} \cdot \| \mathcal{P}_{T^{\perp}}(\mathbf{H})\|_{*} \nonumber\\
&- \underbrace{\| \mathcal{P}_{T}\big(\mathbf{G} - \mathbf{W}_{L}^{\mathsf{H}} \mathbf{U}\mathbf{V}^{\mathsf{H}} \mathbf{W}_{R} \big) \|_{\rm F}}_{\leq \frac{1}{5 \|\mathcal{Q}_{\Omega}\|}} \cdot \|\mathcal{P}_{T}(\mathbf{H})\|_{\rm F} \nonumber\\
=& f(\mathbf{Y}) + \tfrac{\| \mathcal{P}_{T^{\perp}}{(\mathbf{H})} \|_{*} }{2}  - \tfrac{\|\mathcal{P}_{T}(\mathbf{H})\|_{\rm F}}{5\|\mathcal{Q}_{\Omega}\|}. \label{eq:longLemma1}
\end{align}
To show that $f(\widehat{\mathbf{M}}) > f(\mathbf{Y})$ holds with strict inequality for $\mathbf{H}\neq 0$, we shall show below that 
\begin{align}\label{eq:ineqClaim}
\| \mathcal{P}_{T^{\perp}}{(\mathbf{H})} \|_{*} \geq \tfrac{1}{2\|\mathcal{Q}_{\Omega}\|} \|\mathcal{P}_{T}(\mathbf{H})\|_{\rm F}.
\end{align}
Thus,
\begin{align}
\tfrac{\| \mathcal{P}_{T^{\perp}}{(\mathbf{H})} \|_{*} }{2}  - \tfrac{\|\mathcal{P}_{T}(\mathbf{H})\|_{\rm F}}{5\|\mathcal{Q}_{\Omega}\|} 
\geq \tfrac{\|\mathcal{P}_{T}(\mathbf{H})\|_{\rm F}}{20\|\mathcal{Q}_{\Omega}\|}.
\end{align}
If $\mathcal{P}_{T}(\mathbf{H}) \neq 0$, the strict inequality in $f(\widehat{\mathbf{M}}) > f(\mathbf{Y})$ is automatically satisfied. Using \eqref{eq:longLemma1}, if $\mathcal{P}_{T}(\mathbf{H}) = 0$ and $\mathcal{P}_{T^{\perp}}{(\mathbf{H})} \neq 0$, again the strict inequality holds. Thus, the equality happens when $\mathcal{P}_{T}(\mathbf{H}) = \mathcal{P}_{T^{\perp}}{(\mathbf{H})} = 0$, i.e., $\mathbf{H} = \mathcal{P}_{T}(\mathbf{H}) + \mathcal{P}_{T^{\perp}}{(\mathbf{H})} = 0$.

To prove \eqref{eq:ineqClaim}, note that
\begin{align}
0 = \mathcal{Q}_{\Omega} (\mathbf{H}) = \mathcal{Q}_{\Omega} \mathcal{P}_{T}(\mathbf{H}) + \mathcal{Q}_{\Omega} \mathcal{P}_{T^{\perp}}(\mathbf{H}),
\end{align}
we know that 
\begin{align}
\|\mathcal{Q}_{\Omega} \mathcal{P}_{T}(\mathbf{H}) \|_{\rm F} = \|\mathcal{Q}_{\Omega} \mathcal{P}_{T^{\perp}}(\mathbf{H})\|_{\rm F}.
\end{align}
On one hand we have,
\begin{align}
&\|\mathcal{Q}_{\Omega} \mathcal{P}_{T}(\mathbf{H}) \|_{\rm F}^2  
= \langle \mathcal{Q}_{\Omega} \mathcal{P}_{T}(\mathbf{H}) \,,\, \mathcal{Q}_{\Omega} \mathcal{P}_{T}(\mathbf{H}) \rangle \nonumber\\
&= \langle  \mathcal{P}_{T}(\mathbf{H}) \,,\, \mathcal{Q}_{\Omega} \mathcal{P}_{T}(\mathbf{H}) \rangle
= \langle  \mathcal{P}_{T}(\mathbf{H}) \,,\,  \mathcal{P}_{T}\mathcal{Q}_{\Omega} \mathcal{P}_{T}(\mathbf{H}) \rangle
\nonumber\\
&=\left\langle \mathcal{P}_{T}(\mathbf{H}),  \mathcal{P}_{T}(\mathbf{H}) \right\rangle - \left\langle \mathcal{P}_{T}(\mathbf{H}), (\mathcal{P}_{T} - \mathcal{P}_{T} \mathcal{Q} _{\Omega}\mathcal{P}_{T}) \mathcal{P}_{T}(\mathbf{H}) \right\rangle \nonumber\\
& \geq \underbrace{(1 - \| \mathcal{P}_{T} - \mathcal{P}_{T} \mathcal{Q} _{\Omega}\mathcal{P}_{T} \|)}_{\geq \tfrac{1}{2}} \| \mathcal{P}_{T}(\mathbf{H}) \|_{\rm F}^{2} \geq \tfrac{1}{2} \| \mathcal{P}_{T}(\mathbf{H}) \|_{\rm F}^{2}.
 \label{case2_2_1}
	\end{align}
On the other hand,
\begin{align}
	\|  \mathcal{Q}_{\Omega} \mathcal{P}_{T}(\mathbf{H}) \|_{\rm F} 
	=\|  \mathcal{Q}_{\Omega} \mathcal{P}_{T^{\perp}}(\mathbf{H}) \|_{\rm F} \leq \| \mathcal{Q}_{\Omega}  \| \cdot \| \mathcal{P}_{T^{\perp}}(\mathbf{H}) \|_{\rm F},
\end{align}
which in combination with \eqref{case2_2_1} shows that
\begin{align}
\label{eq:PtNormPtperb}
\tfrac{1}{2} \| \mathcal{P}_{T}(\mathbf{H}) \|_{\rm F}^{2} \leq \| \mathcal{Q}_{\Omega}  \| \cdot \| \mathcal{P}_{T^{\perp}}(\mathbf{H}) \|_{\rm F}.
\end{align}
Since for all $\mathbf{M}$ we have $\|\mathbf{M}\|_{\rm F} \leq \|\mathbf{M}\|_{*}$, we conclude that
\begin{align}\label{eq:Nuclear_F_F}
\| \mathcal{P}_{T^{\perp}}(\mathbf{H}) \|_{*} &\geq \| \mathcal{P}_{T^{\perp}}(\mathbf{H}) \|_{\rm F} 
\geq \frac{1}{2 \|\mathcal{Q}_{\Omega}\|} \| \mathcal{P}_{T}(\mathbf{H}) \|_{\rm F} .
\end{align}


\section{ Prof of Lemma 2} 
Let us define the following family of operators  $\mathcal{Z}_{n}: \mathbb{C}^{d_1 \times d_2} \mapsto \mathbb{C}^{d_1 \times d_2} $ as
\begin{align*}
\mathcal{Z}_{n} := \big(\tfrac{\delta_{n}}{p_{n}} - 1\big)\mathcal{P}_T \mathcal{A}_{n} \mathcal{P}_T  \quad \forall ~ n \in [N].
\end{align*}
First, we bound the projection of each $ \mathbf{A}_{n}$ onto the subspace $T$ by using Definition (4).
\begin{align}
  \label{tangent_bound}
\| \mathcal{P}_T(\mathbf{A}_{n}) \|_{\rm F}^2  \leq \| \mathcal{P}_U(\mathbf{A}_{n}) \|_{\rm F}^2 + \| \mathcal{P}_V(\mathbf{A}_{n}) \|_{\rm F}^2 \leq \color{black}\frac{2{\tilde{\mu}_{n}}\tilde{K}} {N}.
\end{align}
Then, we can check that for any $ \mathbf{M} \in \mathbb{C}^{d_1 \times d_2} $ we have
\begin{equation}\label{eq:ZF}
\begin{aligned}
\| \mathcal{Z}_{n} (\mathbf{M}) \|_{\rm F} = & \| \big(\tfrac{\delta_{n}}{p_{n}} - 1\big) \underbrace{\langle  \mathbf{A}_{n}, \mathcal{P}_T(\mathbf{M}) \rangle}_{=\langle  \mathcal{P}_T(\mathbf{A}_{n}), \mathbf{M} \rangle} \mathcal{P}_T(\mathbf{A}_{n})  \|_{\rm F}\\
& \leq \tfrac{1}{p_{n}} \|  \mathcal{P}_T (\mathbf{A}_{n}) \|_{\rm F}^2 \| \mathbf{M} \|_{\rm F}.
\end{aligned}
\end{equation}
Therefore, the operator norm $ \| \mathcal{Z}_{n} \| $ is upper-bounded as
\begin{equation}
\label{spectral_abs_bound}
\begin{aligned}
\| \mathcal{Z}_{n} \| \leq \frac{1}{p_{n}} \|  \mathcal{P}_T (\mathbf{A}_{n}) \|_{\rm F}^2  
\leq  { \frac{2}{p_{n}} \frac{\tilde{\mu}_{n}\tilde{K}}{N} }  \leq \frac{2}{c_0 \log(N)},
\end{aligned}
\end{equation}
where we used \eqref{tangent_bound} and $ p_{n} \geq c_0 \frac{\tilde{\mu}_{n}\tilde{K} \log(N) } {N} $  in the last two inequality, respectively. 

It is not difficult to see that $\sum_{n =1}^{n}\mathcal{Z}_{n} = \mathcal{P}_T \mathcal{Q}_{\Omega} \mathcal{P}_T - \mathcal{P}_T $. 
Since $ \mathbb{E}[\mathcal{Q}_{\Omega} ] = \mathcal{I} $, the latter result shows that $ \mathbb{E}[\mathcal{Z}_{n}] = \mathbf{0}$.  
Besides, for any $\mathbf{M} \in \mathbb{C}^{d_1 \times d_2}$, if $\mathcal{Z}^2_{n}(\mathbf{M})$ represents $\mathcal{Z}_{n}^{*}\big(\mathcal{Z}_{n}(\mathbf{M})\big)$, then, we have
\begin{align}\label{eq:EZ2}
& \Big\| \sum_{n} { \mathbb{E}[ \mathcal{Z}^2_{n}(\mathbf{M}) ] } \Big\|_{\rm F} = \big\| \sum_{n} \mathbb{E}\big[(\tfrac{\delta_{n}}{p_{n} }-1)^2\big] \langle \mathbf{A}_{n}, \mathcal{P}_T(\mathbf{M}) \rangle \nonumber\\
& \times\langle \mathbf{A}_{n}, \mathcal{P}_T (\mathbf{A}_{n}) \rangle \mathcal{P}_T (\mathbf{A}_{n}) \big\|_{\rm F} \nonumber\\
& \leq \max_{n} \frac{1-p_{n}}{p_{n}} \| \mathcal{P}_T (\mathbf{A}_{n}) \|_{\rm F}^2 \big\|  \sum_{n}{ \langle \mathbf{A}_{n}, \mathcal{P}_T(\mathbf{M}) \rangle \mathcal{P}_T(\mathbf{A}_{n}) }   \big\|_{\rm F} \nonumber\\
& \leq  \max_{n} \frac{1}{p_{n}} \|  \mathcal{P}_T (\mathbf{A}_{n}) \|_{\rm F}^2 \| \mathbf{M} \|_{\rm F}.
\end{align}
Therefore, similar to \eqref{eq:EZ2} the operator norm $ \Big\|\sum_{n} { \mathbb{E}[ \mathcal{Z}_{n}^2 ] }\Big\|$ is upper-bounded as
\begin{align*}
\Big\|\sum_{n} { \mathbb{E}[ \mathcal{Z}_{n}^2 ] }\Big\| \leq \max_{n}\frac{1}{p_{n}} \|  \mathcal{P}_T (\mathbf{A}_{n}) \|_{\rm F}^2 \leq \frac{2}{c_0\log(N)}
\end{align*}

Now, by the matrix Bernstein inequality, for $c_0 \geq \frac{56}{3} (b_1 + 1)$, we know the existence of some constant $ 0 < \epsilon \leq \frac{1}{2} $ such that
\begin{align}
\label{lemma_1}
\big\| \sum_{n} \mathcal{\mathcal{Z}}_{n} \big\| = \| \mathcal{P}_{T} - \mathcal{P}_{T} \mathcal{Q} _{\Omega}\mathcal{P}_{T}  \| \leq \epsilon,
\end{align}
with a probability exceeding $ 1 - N^{- b_1} $.


\section{ Prof of Lemma \ref{lemma4}} 
\label{lem4_proof}
By defining $\mathbf{S}_{n} \in \mathbb{C}^{{d}_1 \times {d}_2} $ as
\begin{align}\label{eq:Sdef}
\mathbf{S}_{n} :=   \big(\tfrac{\delta_{n}}{p_{n}}-1\big)  \mathcal{A}_{n} (\mathbf{M}) , \quad n \in [N] 
\end{align}
we have $\mathcal{Q}_{\Omega} - \mathcal{I} = \mathcal{A}_{\Omega}-\mathcal{A}=\sum_{n\in [N]}\mathbf{S}_{n}$. Besides, $ \mathbb{E}[\mathbf{S}_{n}] = 0 $. 
We first bound $\| \mathbf{S}_{n} \|$ as
\begin{equation}
\begin{aligned}
\| \mathbf{S}_{n} \| &= \big\| \big(\tfrac{\delta_{n}}{p_{n}}-1\big)  \big\langle \mathbf{A}_{n} , \mathbf{M} \big\rangle \mathbf{A}_{n}  \big\| 
\leq  {\tfrac{1}{p_{n}}  \Big| \tfrac{\left\langle  \mathbf{A}_{n}, \mathbf{M} \right\rangle }{\sqrt{\omega_{n}}} \Big|} \\
& \leq \tfrac{1}{c_0 \tilde{K} \mathsf{R}_{\mathscr{L}} \log(N)}  \Big| \tfrac{\left\langle  \mathbf{A}_{n}, \mathbf{M} \right\rangle N}{\sqrt{\omega_{n}} \tilde{\mu}_{n} \tilde{K}} \Big|  ,
\end{aligned}
\end{equation}
{\change  where we used $|\frac{\delta_{n}}{p_{n}}-1 | \leq \frac{1}{p_n}$, $\|\mathbf{A}_{n}\| = \frac{1}{\sqrt{\omega_n}}$ and $ p_{n} \geq c_0 \frac{\tilde{\mu}_{n} \mathsf{R}_{\mathscr{L}} \log(N) \tilde{K}^2}{N} $. Therefore,
\begin{align}\label{eq:S_norm}
\max_n \| \mathbf{S}_{n} \| \leq \tfrac{\| \mathbf{M} \|_{\rm \mathcal{A}, \infty}}{c_0 \mathsf{R}_{\mathscr{L}} \tilde{K}\log(N)}.
\end{align}
Recalling the definition of $\mathbf{S}_{n}$ in \eqref{eq:Sdef} and the fact that $\mathcal{A}_{n} (\mathbf{M}) = \langle \mathbf{A}_{n} , \mathbf{M} \rangle \mathbf{A}_{n}$ we can check that
\begin{align}
\mathbf{S}_{n} \mathbf{S}_{n}^{\mathsf{H}} = (\tfrac{\delta_{n}}{p_{n}}-1)^2 \left| \left\langle \mathbf{A}_{n} , \mathbf{M} \right\rangle  \right|^2 \mathbf{A}_{n} \mathbf{A}_{n}^{\mathsf{T}}.
\end{align}
This enables us to write 
\begin{align}\label{eq:SSH}
\Big\|&\mathbb{E} \Big[ \sum_{n} \mathbf{S}_{n} \mathbf{S}_{n}^{\mathsf{H}} \Big] \Big\| =\Big\| \sum_{n} \underbrace{\mathbb{E}  \big[(\tfrac{\delta_{n}}{p_{n}} -1)^2\big] }_{\frac{1-p_n}{p_n}} \left| \left\langle \mathbf{A}_{n} , \mathbf{M} \right\rangle  \right|^2 \mathbf{A}_{n} \mathbf{A}_{n}^{\mathsf{T}} \Big\| \nonumber\\
&\leq \sum_{n} \frac{1}{p_n} \left| \left\langle \mathbf{A}_{n} , \mathbf{M} \right\rangle  \right|^2 \|\mathbf{A}_{n} \mathbf{A}_{n}^{\mathsf{T}}\| \leq \sum_{n} \tfrac{1}{p_n} \tfrac{\left| \left\langle \mathbf{A}_{n} , \mathbf{M} \right\rangle  \right|^2}{\omega_n} .
\end{align}
Since $\|\mathbf{A}_{n}^{\mathsf{T}} \mathbf{A}_{n}\| =\|\mathbf{A}_{n} \mathbf{A}_{n}^{\mathsf{T}}\|=\frac{1}{\omega_n}$, with a similar technique we can obtain
\begin{align}\label{eq:SHS}
\Big\|\mathbb{E}  \Big[ \sum_{n} \mathbf{S}_{n}^{\mathsf{H}} \mathbf{S}_{n} \Big] \Big\|
\leq \sum_{n} \tfrac{1}{p_n} \tfrac{\left| \left\langle \mathbf{A}_{n} , \mathbf{M} \right\rangle  \right|^2}{\omega_n} .
\end{align}
Now, by using $p_{n} \geq c_0 \frac{\tilde{\mu}_{n} \mathsf{R}_{\mathscr{L}} \log(N) \tilde{K}^2}{N} $ in \eqref{eq:SSH} and \eqref{eq:SHS}, we derive
\begin{align}\label{eq:SSH_SHS}
\max &\Big(\Big\|\mathbb{E}  \Big[ \sum_{n} \mathbf{S}_{n}^{\mathsf{H}} \mathbf{S}_{n} \Big] \Big\| \,,\, \Big\|\mathbb{E}  \Big[ \sum_{n} \mathbf{S}_{n} \mathbf{S}_{n}^{\mathsf{H}} \Big] \Big\|\Big) \nonumber\\
&\leq \sum_{n} \tfrac{1}{p_n} \tfrac{\left| \left\langle \mathbf{A}_{n} , \mathbf{M} \right\rangle  \right|^2}{\omega_n} \leq \tfrac{1}{c_0\tilde{K} \mathsf{R}_{\mathscr{L}} \log(N)} \| \mathbf{M} \|_{\rm \mathcal{A}, 2}^2.
\end{align}}
Finally, by combining \eqref{eq:S_norm} and \eqref{eq:SSH_SHS}  with the Bernstein inequality (\cite[Appendix~A]{chen2014robust}), the lemma can be concluded.


\section{ Prof of Lemma \ref{lemma5}} 
\label{lem2_proof}

If we define the vector $ \mathbf{z}_{\alpha} \in \mathbb{C}^{n} $ as 
\begin{align}
\label{def_z_k_l}
\mathbf{z}_{\alpha}({j}) := \sqrt{\tfrac{N}{\tilde{K}\mu_{j} \omega_{j} }} \Big\langle \mathbf{A}_{j} , \big(\tfrac{\delta_{\alpha}}{p_{\alpha}} \mathcal{P}_T \mathcal{A}_{\alpha} - \mathcal{P}_T \mathcal{A}_{\alpha}\big)\mathbf{M} \Big\rangle,
\end{align}
for $ j \in [N] $, due to the  definition of $ \| \mathbf{M} \|_{\rm \mathcal{A},2} $ we have that
$\| (   {\mathcal{P}_T \mathcal{Q}_{\Omega}} - \mathcal{P}_T  )  (\mathbf{M}) \|_{\rm \mathcal{A}, 2} = \left\| \sum_{\alpha} { \mathbf{z}_{\alpha} } \right\|_{\rm 2}$. 
First, we bound $\left\| \mathbf{z}_{\alpha} \right\|_{\rm 2}$ and $\Big\| \mathbb{E}\Big[\sum_{\alpha}{\mathbf{z}_{\alpha} \mathbf{z}_{\alpha}^*}\Big] \Big\| $; next, we use the Bernstein inequality to control $\left\| \sum_{\alpha} { \mathbf{z}_{\alpha} } \right\|_{\rm 2}$. It is not difficult to see $ \mathbb{E}\left[ \mathbf{z}_{\alpha} \right] = \mathbf{0}  $. We bound $\left\| \mathbf{z}_{\alpha} \right\|_{\rm 2}$ as  {\change
\begin{align}
\nonumber
&\left\| \mathbf{z}_{\alpha} \right\|_{\rm 2} = \sqrt{\sum_{j} \tfrac{N}{\tilde{K} \mu_{j} \omega_{j}}  
	\big|  \big\langle  \mathbf{A}_{j}, (\tfrac{\delta_{\alpha}}{p_{\alpha}} - 1) \mathcal{P}_T \underbrace{\mathcal{A}_{\alpha} \left( \mathbf{M} \right)}_{\langle \mathbf{A}_{\alpha}, \mathbf{M}\rangle \mathbf{A}_{\alpha}} \big\rangle  \big|^2 } \\ \nonumber
& \leq \big|\tfrac{\delta_{\alpha}}{p_{\alpha}} - 1\big| \left| \left\langle \mathbf{A}_{\alpha}, \mathbf{M} \right\rangle \right| \sqrt{\sum_{j} \tfrac{N}{ \tilde{K} \mu_{j} \omega_{j}} \left| \left\langle  \mathbf{A}_{j}, \mathcal{P}_T \left( \mathbf{A}_{\alpha} \right) \right\rangle  \right|^2}\\ \nonumber
& \leq \big|\tfrac{\delta_{\alpha}}{p_{\alpha}} - 1\big| \left| \left\langle \mathbf{A}_{\alpha}, \mathbf{M} \right\rangle \right| \sqrt{\sum_{j} \tfrac{N}{\tilde{K} \mu_{j} \omega_{j}} \underbrace{\| \mathcal{P}_T \left( \mathbf{A}_{j} \right) \|_{\rm F}^2}_{\leq \tfrac{2\mu_j\tilde{K}}{N} } 
\underbrace{\| \mathcal{P}_T \left( \mathbf{A}_{\alpha} \right) \|_{\rm F}^2}_{ \leq \tfrac{2\mu_{\alpha}\tilde{K}}{N} } }
\\\label{lemma5_bound_1}
& = 2 \big|\tfrac{\delta_{\alpha}}{p_{\alpha}} - 1\big| \left| \left\langle \mathbf{A}_{\alpha}, \mathbf{M} \right\rangle \right| \sqrt{ \frac{\mu_{\alpha}\tilde{K}}{N} \underbrace{\sum_{j} \frac{1}{\omega_{j}}}_{ \leq \mathsf{R}_{\mathscr{L}}}  } .
\end{align}
 We recall that $ \mathsf{R}_{\mathscr{L}} = \sum_{j \in [N]}  \big\|\mathbf{A}_{j} \odot \mathbf{A}_{j}\big\|_{\infty \rightarrow \infty}$}.
%
Moreover, $ \frac{\mu_{\alpha}\tilde{K}}{N} = \frac{1}{\omega_{\alpha}}\sum_{i=1}^{\omega_{\alpha}} \| \mathbf{U}\mathbf{e}_{i}^{{d}_1} \|_2^2  \leq \frac{1}{\omega_{\alpha}} \| \mathbf{U}\|_{\rm F}^{2} = \frac{\tilde{K}}{\omega_{\alpha}} $. Hence,
\begin{align}
\label{lemma5_norm_max_A}
  \left\| \mathbf{z}_{\alpha} \right\|_{\rm 2} & \leq \underbrace{\big|\tfrac{\delta_{\alpha}}{p_{\alpha}} - 1\big|}_{\leq \frac{1}{p_n}} \tfrac{2}{ \sqrt{\omega_{\alpha}}} \left| \left\langle \mathbf{A}_{\alpha}, \mathbf{M} \right\rangle \right| \sqrt{\mathsf{R}_{\mathscr{L}} \tilde{K} } \\
& \leq \tfrac{2}{c_0 \log(N)} \frac{N \left| \left\langle \mathbf{A}_{\alpha}, \mathbf{M} \right\rangle \right| }{\mu_{\alpha}  \tilde{K} \sqrt{\omega_{\alpha}}}  = \tfrac{2}{c_0 \log(N)} \left\| \mathbf{M} \right\| _{\rm  \mathcal{A},\infty}, \label{lemma5_norm_max}
\end{align}
where $ p_{n} \geq c_0 \frac{\tilde{\mu}_{n} \mathsf{R}_{\mathscr{L}} \tilde{K}^2 \log(N)}{N} $ and $ \sqrt{\mathsf{R}_{\mathscr{L}} \tilde{K}} \geq 1 $. 
Moreover,  we bound $\Big\| \mathbb{E}\Big[\sum_{\alpha}{\mathbf{z}_{\alpha} \mathbf{z}_{\alpha}^*}\Big] \Big\|$ as $ \Big\| \mathbb{E}\Big[\sum_{\alpha}{\mathbf{z}_{\alpha} \mathbf{z}_{\alpha}^*}\Big] \Big\|  \leq \sum_{\alpha} \mathbb{E}\big[ \|\mathbf{z}_{\alpha}\|_2^2 \big]$ We can now use \eqref{lemma5_norm_max_A} to conclude
\begin{align}
\nonumber
\Big\| \mathbb{E}&\Big[\sum_{\alpha}{\mathbf{z}_{\alpha} \mathbf{z}_{\alpha}^*}\Big] \Big\|
= \sum_{\alpha} { 4 \mathbb{E}\Big[\big( \tfrac{\delta_{\alpha}}{p_{\alpha}} - 1 \big)^2\Big] \tfrac{\left| \left\langle  \mathbf{A}_{\alpha}, \mathbf{M} \right\rangle  \right|^2}{\omega_{\alpha}} } \tilde{K} \mathsf{R}_{\mathscr{L}} \\ \nonumber
& \leq \tfrac{4}{p_{\alpha}} \tilde{K} \mathsf{R}_{\mathscr{L}} \sum_{\alpha} \tfrac{\left| \left\langle  \mathbf{A}_{\alpha}, \mathbf{M} \right\rangle  \right|^2}{\omega_{\alpha}}\leq \tfrac{4N}{c_0 \mu_{\alpha} \tilde{K} \log(N)} \sum_{\alpha} \tfrac{\left| \left\langle  \mathbf{A}_{\alpha}, \mathbf{M} \right\rangle  \right|^2}{\omega_{\alpha}} \\
&= \tfrac{4}{c_0 \log\left( N \right)} \left\| \mathbf{M} \right\| _{\rm \mathcal{A},2 }^2,
\end{align}
where the last inequality comes from \eqref{def_norm_2} and previous one from  $ p_{\alpha} \geq c_0 \frac{{\mu}_{\alpha} \tilde{K}^2 \mathsf{R}_{\mathscr{L}} \log(N)}{N} $. In addition, $\Big\| \mathbb{E}\Big[\sum_{\alpha}{\mathbf{z}_{\alpha} \mathbf{z}_{\alpha}^*}\Big] \Big\| = \Big\| \mathbb{E}\Big[\sum_{\alpha}{\mathbf{z}_{\alpha}^* \mathbf{z}_{\alpha}}\Big] \Big\|.$ One can conclude the proof by applying \cite[Appendix~A]{chen2014robust}.

\section{Prof of Lemma \ref{lemma_6} } 
\label{lem6_proof}

Lets define  
\begin{equation*}
\begin{aligned}
\mathbf{z}_{\alpha}({n}) := \tfrac{N}{\tilde{K}\tilde{\mu}_{n} \sqrt{\omega_{n}} } \Big\langle \mathbf{A}_{n} , \big(\tfrac{\delta_{\alpha}}{p_{\alpha}} \mathcal{P}_T \mathcal{A}_{\alpha} - \mathcal{P}_T \mathcal{A}_{\alpha}\big) \mathbf{M} \Big\rangle,
\end{aligned}
\end{equation*}
for $ n \in [N] $, we have that
$$\| ( {\mathcal{P}_T \mathcal{Q}_{\Omega}} - \mathcal{P}_T  )  (\mathbf{M}) \|_{\rm \mathcal{A}, \infty} = \Big\| \sum_{\alpha} { \mathbf{z}_{\alpha} } \Big\|_{\infty} = \max_{n} \Big| \sum_{\alpha} { \mathbf{z}_{\alpha}(n) } \Big|.$$

To bound $\| ( {\mathcal{P}_T \mathcal{Q}_{\Omega}} - \mathcal{P}_T  )  (\mathbf{M}) \|_{\rm \mathcal{A}, \infty} $, we focus on bounding $|\mathbf{z}_{\alpha}(n)|$s, which in turn requires bounding  $\left| \left\langle \mathbf{A}_{\alpha}, \mathcal{P}_{T} (\mathbf{A}_{n})  \right\rangle \right|$:
%
\begin{align}
\nonumber
\left| \left\langle \mathbf{A}_{\alpha}, \mathcal{P}_{T}( \mathbf{A}_{n} ) \right\rangle \right| &\leq  \left| \left\langle \mathbf{A}_{\alpha}, \mathcal{P}_{U}(\mathbf{A}_{n})  \right\rangle \right|  + \big| \big\langle \mathbf{A}_{\alpha}, \mathcal{P}_{V}(\mathbf{A}_{n}) \big\rangle \big|  \\ \label{eq:PTAk}
& + \big| \big\langle \mathbf{A}_{\alpha}, \mathcal{P}_{U} \mathcal{P}_{V} (\mathbf{A}_{n}) \big\rangle \big|,
\end{align}
in which we used the definition of $\mathcal{P}_{T}$ in (A.7). 
Next, we upper-bound  each term separately:
\begin{align} 
| &\left\langle \mathbf{A}_{\alpha}, \mathcal{P}_{U}( \mathbf{A}_{n} ) \right\rangle |
= \left| \left\langle \mathcal{P}_{U}(\mathbf{A}_{\alpha}), \mathcal{P}_{U}( \mathbf{A}_{n} ) \right\rangle \right| \nonumber \\
&\leq  \|\mathcal{P}_{U}(\mathbf{A}_{\alpha})\|_{\rm F} \cdot \|\mathcal{P}_{U}(\mathbf{A}_{n})\|_{\rm F}  \\
& \leq \sqrt{\tfrac{\tilde{K} \mu_{\alpha}}{N}} \sqrt{\tfrac{\tilde{K} \tilde{\mu}_{n}}{N}} = \tfrac{\tilde{K}\sqrt{\tilde{\mu}_{n}\mu_{\alpha}}}{N} \nonumber
\end{align}
Recalling the definition of $\mathcal{P}_{U}$ in (18a), we know that $\|\mathcal{P}_{U}(\sqrt{\omega_i}\mathbf{A}_{i})\|_2^2$ is the sum of the squared $\ell_2$-norm of the some rows of $\mathbf{W}_{L}^{\mathsf{H}} \mathbf{U} \left( \mathbf{{U}}^{\mathsf{H}} \mathbf{W}_{L} \mathbf{W}_{L}^{\mathsf{H}}\mathbf{{U}} \right)^{-1} \mathbf{{U}}^{\mathsf{H}}\mathbf{W}_{L}$ depending on lifting operator.
{ \change Therefore, 
\begin{align}
 \|\mathcal{P}_{U}(\sqrt{\omega_i}\mathbf{A}_{i})\|_2^2 &\leq \nonumber
 \| \mathbf{W}_{L}^{\mathsf{H}} \mathbf{U} \left( \mathbf{{U}}^{\mathsf{H}} \mathbf{W}_{L} \mathbf{W}_{L}^{\mathsf{H}} \mathbf{{U}} \right)^{-1} \mathbf{{U}}^{\mathsf{H}} \mathbf{W}_{L}\|_{2}^2\\
 &= \tilde{K}.
\end{align}
Similarly, we have 
\begin{align}
\|\mathcal{P}_{V}(\sqrt{\omega_i}\mathbf{A}_{i})\|_2^2 \leq \tilde{K}.
\end{align}
As a result, $\mu_{i}\omega_{i}\leq N$ for all $i\in [N]$. This implies that $\mu_{\alpha} \leq \frac{N}{\omega_{\alpha}}$, and }
%
\begin{align}
\left| \left\langle \mathbf{A}_{\alpha}, \mathcal{P}_{U}( \mathbf{A}_{n} ) \right\rangle \right| & \leq \tilde{K} \sqrt{\tfrac{\tilde{\mu}_{n}}{\omega_{\alpha}N}}. \label{eq:UUAk}
\end{align}
With a similar argument, we have 
\begin{align}
\big| \big\langle \mathbf{A}_{\alpha},  \mathcal{P}_{V}( \mathbf{A}_{n}) \big\rangle \big| \leq 
 \tilde{K}\sqrt{\tfrac{\tilde{\mu}_{n}}{\omega_{\alpha}N}}.
\label{eq:VVAk}
\end{align}
The last term in \eqref{eq:PTAk} to bound is $ | \langle \mathbf{A}_{\alpha},  \mathcal{P}_{U}\mathcal{P}_{V}(\mathbf{A}_{n}) \rangle |$ for which we write:
%
\begin{align}
\nonumber
 |& \langle \mathbf{A}_{\alpha},  \mathcal{P}_{U}\mathcal{P}_{V}(\mathbf{A}_{n}) \rangle | = | \langle \mathcal{P}_{U}(\mathbf{A}_{\alpha}),  \mathcal{P}_{V}(\mathbf{A}_{n}) \rangle | \\
&\leq  \|\mathcal{P}_{U}(\mathbf{A}_{\alpha})\|_{\rm F} \cdot \|\mathcal{P}_{V}(\mathbf{A}_{n})\|_{\rm F} \leq \sqrt{\tfrac{\tilde{K} \mu_{\alpha}}{N}} \sqrt{\tfrac{\tilde{K} \tilde{\mu}_{n}}{N}}  \leq  \tilde{K}\sqrt{\tfrac{\tilde{\mu}_{n}}{\omega_{\alpha}N}}.
\label{eq:UUAkVV}
\end{align}
where the first inequality comes from the Cauchy-Schwartz inequality. Substituting \eqref{eq:UUAk}, \eqref{eq:VVAk}, and \eqref{eq:UUAkVV} into \eqref{eq:PTAk}, we obtain
\begin{align}
\label{base_matrix_bound}
\left| \left\langle \mathbf{A}_{\alpha}, \mathcal{P}_{T}( \mathbf{A}_{n} ) \right\rangle \right| \leq 3 \tilde{K}\sqrt{\tfrac{\tilde{\mu}_{n}}{\omega_{\alpha}N}}.
\end{align} 
On one hand, we can bound $ \left| \mathbf{z}_{\alpha}  \right| $ as 
\begin{align} 
\nonumber
| &\mathbf{z}_{\alpha}(n)| =  \tfrac{N}{\tilde{\mu}_{n}\tilde{K}}\sqrt{ \tfrac{1}{\omega_{n}} }  
\big| \big( \tfrac{\delta_{\alpha}}{p_{\alpha}} -1 \big) \left\langle \mathbf{A}_{\alpha}, \mathbf{M} \right\rangle \left\langle \mathbf{A}_{\alpha}, \mathcal{P}_T \left( \mathbf{A}_{n} \right)  \right\rangle  \big| \\\nonumber
& \leq \tfrac{N}{\tilde{\mu}_{n}\tilde{K}}\sqrt{ \tfrac{1}{\omega_{n}} } \tfrac{1}{p_{\alpha}} \left| \left\langle \mathbf{A}_{\alpha}, \mathbf{M} \right\rangle \right| \left| \left\langle \mathbf{A}_{\alpha}, \mathcal{P}_T \left( \mathbf{A}_{n} \right)  \right\rangle \right| \\ \label{lemma4_abs_1}
& \leq \tfrac{N}{\tilde{\mu}_{n}\tilde{K}}\sqrt{ \tfrac{1}{\omega_{n}} } \tfrac{\left| \left\langle \mathbf{A}_{\alpha}, \mathbf{M} \right\rangle \right|}{p_{\alpha}} 3\tilde{K} \sqrt{\tfrac{\tilde{\mu}_{n}}{\omega_{\alpha}N}}  = \tfrac{3\left| \left\langle \mathbf{A}_{\alpha}, \mathbf{M} \right\rangle \right|}{\sqrt{\omega_{\alpha}}p_{\alpha}} \sqrt{\tfrac{N}{\omega_{n}\tilde{\mu}_{n}}}, 
\end{align}
where we used \eqref{base_matrix_bound} to achieve \eqref{lemma4_abs_1}. Invoking {\change \eqref{eq:Condlem6}} and using the fact that   { \change
\begin{align}
\omega_{n}\tilde{\mu}_{n} 
&\geq \frac{N}{\tilde{K}} \min_{i}\{\omega_i\|\mathcal{P}_{U}(\mathbf{A}_{i})\|_{\rm F}^2,\omega_i \|\mathcal{P}_{V}(\mathbf{A}_{i})\|_{\rm F}^2 \},
\end{align}}
we have that $\omega_{n}\tilde{\mu}_{n} \geq \frac{N}{8\tilde{K}\log (N)}$ which leads to   
\begin{align}
\left| \mathbf{z}_{\alpha}(n)  \right| & \leq\tfrac{3\sqrt{8}\left| \left\langle \mathbf{A}_{\alpha}, \mathbf{M} \right\rangle \right|\sqrt{\tilde{K}}\log (N)}{\sqrt{\omega_{\alpha}}p_{\alpha}}  \leq \tfrac{3\sqrt{8}\left\| \mathbf{M} \right\| _{\rm  \mathcal{A},\infty}}{c_0 },
\label{lemma4_abs_2}
\end{align}
{\change by applying $ p_{\alpha} \geq c_0 \frac{{\mu}_{\alpha} \log(N) \tilde{K}^{1.5}}{N} $. On the other hand, similar to \eqref{lemma4_abs_2} one can bound $ \mathbb{E} \big[  \sum_{\alpha} |\mathbf{z}_{\alpha}(n) |^2   \big] $ as
\begin{align}
\nonumber
&  \mathbb{E} \Big[  \sum_{\alpha} |\mathbf{z}_{\alpha}(n) |^2   \Big]  = \sum_{\alpha}  \mathbb{E}\Big( \tfrac{\delta_{\alpha}}{p_{\alpha}} -1 \Big)^2 \times \\\nonumber
& \tfrac{N^2}{\tilde{\mu}_{n}^2\tilde{K}^2}  \tfrac{1}{\omega_{n}} \left| \left\langle \mathbf{A}_{\alpha}, \mathbf{M} \right\rangle \right|^2 \left| \left\langle \mathbf{A}_{\alpha}, \mathcal{P}_T \left( \mathbf{A}_{n} \right)  \right\rangle \right|^2 \\
& \leq \tfrac{72}{c_0 } \sum_{\alpha}\tfrac{\left| \left\langle \mathbf{A}_{\alpha}, \mathbf{M} \right\rangle \right|^2 N}{\tilde{K} {\mu_{\alpha} {\omega_{\alpha}} }} 
= \tfrac{72\left\| \mathbf{M} \right\| _{\rm  \mathcal{A},2 }^2}{c_0},
\end{align}
where $ p_{\alpha} \geq c_0 \frac{{\mu}_{\alpha} \log(N) \tilde{K}^2}{N} $. The Bernstein inequality in \cite[Appendix~A]{chen2014robust} confirms that 
$\left\| \left( {\mathcal{P}_T \mathcal{Q}_{\Omega}} - \mathcal{P}_T  \right) (\mathbf{M}) \right\|_{\rm \mathcal{A},\infty  } 
\leq \sqrt{72} \left(  \sqrt{\tfrac{2(b_4+1)}{c_0}} \left\| \mathbf{M} \right\|_{\rm \mathcal{A}, 2} + \tfrac{2(b_4+1)}{3c_0} \left\| \mathbf{M} \right\|_{\rm \mathcal{A}, \infty } \right)$ with probability at least $1-N^{-b_4+1}$, where $ c_0 \geq 144(b_4 + 1)$.}


\section{Proof of Corollary 1}
From the proofs of Theorem 1, we know that $\tilde{\mu}_{n}$s determine the minimum number of sufficient observations for perfect recovery. Therefore, we upper-bound $\tilde{\mu}_{n}$s based on weight matrices. {\change First, we recall from (17) in the main manuscript that}
\begin{align}
\tilde{\mu}_n = \tfrac{N}{\tilde{K}} \, \max \big( \|\mathcal{P}_{{U}}(\mathbf{A}_{n})\|_{\rm F}^2 \,,\, \|\mathcal{P}_{{V}}(\mathbf{A}_{n})\|_{\rm F}^2 \big).
\end{align}
It follows that
\begin{align}
\label{eq:2}
&\| \mathcal{P}_{{U}}(\mathbf{A}_{n})\|_{\rm F}^2 \leq  \tfrac{\left\| \mathbf{W}_{L} \mathbf{A}_{n} \right\|_{\rm F}^2 \left\| \mathbf{{U}}^{\mathsf{H}} \right\|^2}  {\sigma_{\tilde{K}}^2 \left( \mathbf{W}_{L}^{\rm T}\mathbf{U} \right)},  
\end{align}
where $ \sigma_{\tilde{K}} (.) $ denotes the $\tilde{K}$-th singular value. One can bound $ \mathcal{P}_{V} $ similar to $ \mathcal{P}_{U} $ as
\begin{align}
\| \mathcal{P}_{V}(\mathbf{A}_{n})\|_{\rm F}^2 \leq \tfrac{\left\| \mathbf{A}_{n} \mathbf{W}_{R}^{\mathsf{T}}  \right\|_{\rm F}^2 \left\| \mathbf{V}^{\mathsf{H}} \right\|^2}  {\sigma_{\tilde{K}}^2 \left( \mathbf{W}_{R}^{\mathsf{T}}\mathbf{V} \right)}.  \label{eq:3}
\end{align}
Next, we aim to simplify \eqref{eq:2} and \eqref{eq:3} by using the diagonal assumption of the weight matrices. Since $ \mathbf{W}_{L}^{\mathsf{T}}\mathbf{U} $ is of rank $ \tilde{K} $ and $ \mathbf{W}_{L} $ is diagonal, we can bound $  \sigma_{\tilde{K}} \left( \mathbf{W}_{L}^{\mathsf{T}}\mathbf{U} \right) $ as
\begin{align}
	\sigma^2_{\tilde{K}} \left(\mathbf{W}_{L}^{\mathsf{T}}\mathbf{U} \right) &= \min_{\left\| \mathbf{x} \right\| = 1} {\left\| \mathbf{W}_{L}\mathbf{U}\mathbf{x} \right\|_2^2} = \min_{\left\| \mathbf{x} \right\| = 1} \sum_{k=1}^{{d}_1} { w_{L,k} \left| \mathbf{e}_{k}^{{d}_1\mathsf{T}}\mathbf{U}\mathbf{x} \right|^2}, \label{minimization_1}
\end{align}
where the diagonal elements of $\mathbf{W}_{L}$ are $\sqrt{w_{L,k}}$s. 
Let $ \mathbf{z}_{k} = \left| \mathbf{e}_{k}^{{d}_1{\mathsf{T}}}\mathbf{U}\mathbf{x} \right|^2 $ for every $ k \in [{d}_1] $; alternatively,  $ \mathbf{z}_{k} =\big| \mathbf{u}_k \mathbf{x}\big|^2$ where $\mathbf{u}_k$ is the $k$th row of $\mathbf{U}$. Hence, $ \mathbf{z}_{k} \leq \|\mathbf{u}_k\|_2^2 \leq \|\mathbf{U}^{\mathsf{H}}\|^2$. Besides, $ \left\| \mathbf{W}_{L}^{\mathsf{T}}\mathbf{U}\mathbf{x} \right\|_2^2 = \sum_{k} { w_{L,k}  \mathbf{z}_{k}} $ and 
\begin{align*}
	\sum_{k} {\mathbf{z}_{k}} = \left\| \mathbf{{U}x} \right\|_{\rm F}^2 =  \left\| \mathbf{x} \right\|_{\rm F}^2 = 1.
\end{align*}	
Now we can relax the conditions in \eqref{minimization_1} to get
\begin{align}
\sigma^2_{\tilde{K}} \left(\mathbf{W}_{L}^{\mathsf{T}}\mathbf{U} \right) \geq 
\underset{ z \in \mathbb{R}^{{d}_1}_{+}}{\rm min}& \sum_{k} {  w_{L,k} \mathbf{z}_{k}} \nonumber\\
{\rm st.} &~~ \left\{\begin{array}{c}
0 \leq \mathbf{z}_{k} \leq  \left\| \mathbf{{U}}^{\mathsf{H}} \right\|^2,\\
\sum_{k} {\mathbf{z}_{k}} = 1.
\end{array}
\right.
	\label{eq:1}
\end{align}
The minimization in \eqref{eq:1} is a linear programming problem. By having $w_{L,{1}}\leq \dots \leq w_{L,{{d}_1}}$, it is not difficult to see that the minimizer $\mathbf{z}^*$ of \eqref{eq:1} satisfies
\begin{align}
\mathbf{z}^*_{l_k} = \left\{\begin{array}{ll}
\| \mathbf{U}^{\mathsf{H}} \|^2, & 1\leq k\leq \lfloor \tfrac{1}{\| \mathbf{U}^{\mathsf{H}} \|^2}\rfloor, \\
1- \| \mathbf{U}^{\mathsf{H}} \|^2 \lfloor \tfrac{1}{\| \mathbf{U}^{\mathsf{H}} \|^2}\rfloor, & k=\lfloor \tfrac{1}{\| \mathbf{U}^{\mathsf{H}} \|^2}\rfloor+1,\\
0, & k > \lfloor \tfrac{1}{\| \mathbf{U}^{\mathsf{H}} \|^2}\rfloor+1.
\end{array}
\right.
\end{align}
In addition, the value of the minimizer is given by
\begin{align}
\| \mathbf{U}^{\mathsf{H}} \|^2\sum_{k\in [\nu]}   w_{L,k} &+ 
\big(1- \nu \| \mathbf{U}^{\mathsf{H}} \|^2  \big) w_{L,{\nu+1}} \nonumber\\
&\geq \| \mathbf{U}^{\mathsf{H}} \|^2\sum_{k=\in [\nu]}   w_{L,k}
\end{align}
where $\nu = \lfloor \tfrac{1}{\| \mathbf{U}^{\mathsf{H}} \|^2}\rfloor$. This implies that
\begin{align}
\sigma^2_{\tilde{K}} \left(\mathbf{W}_{L}^{\mathsf{T}}\mathbf{U} \right) \geq 
\| \mathbf{U}^{\mathsf{H}} \|^2\sum_{k=1}^{\big\lfloor \tfrac{1}{\| \mathbf{U}^{\mathsf{H}} \|^2}\big\rfloor}   w_{L,k}.
\end{align}
Now, by substituting this result in \eqref{eq:2} we have
\begin{align}
	\label{eq:UpperPU}
	\| P_{{{U}}}(\mathbf{A}_{n})\|_{\rm F}^2 \leq \tfrac{\left\| \mathbf{W}_{L} \mathbf{A}_{n} \right\|_{\rm F}^2}{ \sum_{i=1}^{\lfloor\frac{N}{\beta \tilde{K}}\rfloor} w_{L,{i}} }
\end{align}
Similarly, for $P_{V}$, we have 
\begin{align}
\label{eq:UpperPV}
\| P_{{V}}(\mathbf{A}_{n})\|_{\rm F}^2 \leq \tfrac{\left\| \mathbf{A}_{n} \mathbf{W}_{R} \right\|_{\rm F}^2}{ \sum_{j=1}^{\lfloor\frac{N}{\beta \tilde{K}}\rfloor}{w_{R,{j}} }}.
\end{align}	
Finally, combining \eqref{eq:UpperPU} and \eqref{eq:UpperPV} leads to
\begin{align}
	\frac{\tilde{\mu}_{n}\tilde{K}}{n} \leq  \max \Big(\tfrac{\left\| \mathbf{A}_{n} \mathbf{{W}_{L}} \right\|_{\rm F}^2}{\sum_{i=1}^{\lfloor\tfrac{N}{\beta \tilde{K}}\rfloor}w_{L,i}}, \tfrac{\left\| \mathbf{A}_{n} \mathbf{{W}_{R}} \right\|_{\rm F}^2}{\sum_{j=1}^{\lfloor\tfrac{N}{\beta \tilde{K}}\rfloor}w_{R,j}}\Big).
\end{align}

\bibliographystyle{elsarticle-num}
\bibliography{IEEEabrv,references}